\documentclass[usegraphicx]{mn2e}

\title[The specific angular momentum of disc galaxies]
      {The specific angular momentum of disc galaxies and its connection with
       galaxy morphology, bar structure and disc gravitational instability}

\author[A. B. Romeo, O. Agertz and F. Renaud]
       {Alessandro B. Romeo,$^{1}$\thanks{E-mail: romeo@chalmers.se}
        Oscar Agertz$^{2}$ and
        Florent Renaud$^{2}$\\
        $^{1}$Department of Space, Earth and Environment,
              Chalmers University of Technology,
              SE-41296 Gothenburg, Sweden\\
        $^{2}$Department of Astronomy and Theoretical Physics,
              Lund University, Box 43,
              SE-22100 Lund, Sweden}

\begin{document}

\date{Accepted 2022 October 20.
      Received 2022 October 10; in original form 2022 April 06}

\pagerange{\pageref{firstpage}--\pageref{lastpage}}

\pubyear{2022}

\maketitle

\label{firstpage}

\begin{abstract}
The specific angular momenta ($j\equiv J/M$) of stars ($j_{\star}$), gas
($j_{\mathrm{gas}}$), baryons as a whole ($j_{\mathrm{b}}$) and dark matter
haloes ($j_{\mathrm{h}}$) contain clues of vital importance about how
galaxies form and evolve.  Using one of the largest samples of disc galaxies
(S0--BCD) with high-quality rotation curves and near-infrared surface
photometry, we perform a detailed comparative analysis of $j$ that stretches
across a variety of galaxy properties.  Our analysis imposes tight
constraints on the `retained' fractions of specific angular momentum
($j_{\star}/j_{\mathrm{h}}$, $j_{\mathrm{HI}}/j_{\mathrm{h}}$ and
$j_{\mathrm{b}}/j_{\mathrm{h}}$), as well as on their systematic trends with
mass fraction and galaxy morphology, thus on how well specific angular
momentum is conserved in the process of disc galaxy formation and evolution.
In particular, one of the most innovative results of our analysis is the
finding that galaxies with larger baryon fractions have also retained larger
fractions of their specific angular momentum.  Furthermore, our analysis
demonstrates how challenging it is to characterize barred galaxies from a
gravitational instability point of view.  This is true not only for the
popular Efstathiou, Lake \& Negroponte bar instability criterion, which fails
to separate barred from non-barred galaxies in about 55\% of the cases, but
also for the mass-weighted Toomre parameter of atomic gas, $\langle
Q_{\mathrm{HI}}\rangle$, which succeeds in separating barred from non-barred
galaxies, but only in a statistical sense.
\end{abstract}


\section{INTRODUCTION}

Specific angular momentum, $j\equiv J/M$, is one of the most fundamental
galaxy properties (see, e.g., Combes 2020; Obreschkow 2020).  Today, four
decades after the pioneering work of Fall (1983), the scaling relation
between stellar specific angular momentum ($j_{\star}$) and stellar mass
($M_{\star}$), $j_{\star}\propto M_{\star}^{s}$ with $s\sim2/3$, has been
confirmed and refined in a wide variety of contexts, not only for nearby
galaxies (e.g., Romanowsky \& Fall 2012; Fall \& Romanowsky 2013, 2018; Lapi
et al.\ 2018; Posti et al.\ 2018) but also for distant galaxies at redshift
$z\la3$ (e.g., Burkert et al.\ 2016; Marasco et al.\ 2019; Sweet et
al.\ 2019; Gillman et al.\ 2020; Bouch\'{e} et al.\ 2021).  A similar scaling
relation has been found for atomic gas, as well as for stars and atomic gas
as a whole: the `baryons' (e.g., Obreschkow \& Glazebrook 2014; Murugeshan et
al.\ 2020; Kurapati et al.\ 2021; Mancera Pi\~{n}a et al.\ 2021a,\,b).
Indeed, all such relations are remarkably similar to the scaling law
$j\propto M^{2/3}$ predicted by tidal torque theory (e.g., Peebles 1969;
Efstathiou \& Jones 1979), which is one of the most fundamental relations for
dark matter haloes (see, e.g., Cimatti et al.\ 2020).

The stellar-to-halo and baryonic-to-halo $j$ ratios,
$j_{\star}/j_{\mathrm{h}}$ and $j_{\mathrm{b}}/j_{\mathrm{h}}$, are of great
theoretical importance because they measure the fractions of specific angular
momentum retained by stars and baryons, i.e.\ how well specific angular
momentum is conserved in the process of galaxy formation and evolution (see,
e.g., Cimatti et al.\ 2020).  To get this information, one needs to know
$j_{\mathrm{h}}$, which is not a truly observable galaxy property.  Soon
after the seminal paper by Romanowsky \& Fall (2012), it has become common
practice to `measure' $j_{\mathrm{h}}$ via the $j_{\mathrm{h}}\propto
M_{\mathrm{h}}^{2/3}$ relation, where $M_{\mathrm{h}}$ can be inferred using
a number of different methods (see, e.g., Wechsler \& Tinker 2018).
Investigations based on this or similar approaches have placed a basic
constraint on $j_{\star}/j_{\mathrm{h}}$ and $j_{\mathrm{b}}/j_{\mathrm{h}}$:
these ratios are typically below unity, the value assumed by classic disc
formation models (e.g., Dutton \& van den Bosch 2012; Romanowsky \& Fall
2012; Kauffmann et al.\ 2015; Lapi et al.\ 2018; Mancera Pi\~{n}a et
al.\ 2021a).  `Biased-collapse' or `inside-out' models of galaxy formation
assume instead that there is a power-law relation between retained fraction
of specific angular momentum and mass fraction, but they do not predict the
actual slopes of the stellar and baryonic relations (e.g., Dutton \& van den
Bosch 2012; Romanowsky \& Fall 2012).  Clearly, $j_{\star}/j_{\mathrm{h}}$
and $j_{\mathrm{b}}/j_{\mathrm{h}}$ are less constrained than the
$j_{\star}\mbox{--}M_{\star}$ and $j_{\mathrm{b}}\mbox{--}M_{\mathrm{b}}$
relations.

Cosmological simulations have struggled to explain the origins of $j_{\star}$
and $j_{\mathrm{b}}$.  Early work found a catastrophic loss of angular
momentum during galaxy assembly, with values of $j_{\star}/j_{\mathrm{h}}$
and $j_{\mathrm{b}}/j_{\mathrm{h}}$ far below those predicted by classic disc
formation models (Navarro \& White 1994; Navarro \& Steinmetz 2000).  This
problem has since been alleviated, thanks to a better understanding of
feedback from massive stars and active galactic nuclei (see, e.g., Naab \&
Ostriker 2017).  Feedback promotes disc formation by preferentially removing
low angular momentum gas from galaxies via outflows (Brook et al.\ 2011;
\"{U}bler et al.\ 2014), as well as by suppressing star formation in the
early Universe, when accreting gas was poor in angular momentum (Agertz \&
Kravtsov 2016; Garrison-Kimmel et al.\ 2018).  Simulations have highlighted
that many mechanisms (e.g.\ inflows, mergers and disc gravitational
instabilities) are responsible for shaping the net angular momentum content
of galaxies, with no consensus on their respective importance (see, e.g.,
Lagos et al.\ 2020).

What makes $j$ a quantity of great astrophysical importance is not only its
relation to the baryonic and dark matter content of galaxies, but also its
connection with galaxy morphology.  This was beautifully illustrated by Fall
(1983).  His fig.\ 1 shows that ellipticals and spirals form two parallel
$j_{\star}\mbox{--}M_{\star}$ tracks, and that ellipticals contain less
$j_{\star}$ on average than spirals of equal $M_{\star}$.  More recent
investigations have generalized these findings to galaxies of various
morphological types, and analysed their trend as a function of bulge mass
fraction, B/T, or as a function of other morphological proxies, not only in
the case of the $j_{\star}\mbox{--}M_{\star}$ relation (e.g., Romanowsky \&
Fall 2012; Cortese et al.\ 2016; Fall \& Romanowsky 2018; Sweet et al.\ 2018)
but also for $j_{\mathrm{b}}$ vs $M_{\mathrm{b}}$ (e.g., Obreschkow \&
Glazebrook 2014; Kurapati et al.\ 2021; Mancera Pi\~{n}a et al.\ 2021a,\,b).
A few investigations also found that early-type galaxies have retained less
specific angular momentum than late-type ones (e.g., Romanowsky \& Fall 2012;
Kauffmann et al.\ 2015), which places another basic constraint on
$j_{\star}/j_{\mathrm{h}}$ and $j_{\mathrm{b}}/j_{\mathrm{h}}$.

$j$ spreads its `tentacles' even deeper, into one of the engines behind the
dynamics of disc galaxies: gravitational instability.  Such a connection was
discovered in the context of bar instability (Christodoulou et al.\ 1995; Mo
et al.\ 1998; van den Bosch 1998), hence it also involves an important aspect
of galaxy morphology: bar structure.  The connection between $j$ and disc
gravitational instability emerges even in the case of local instabilities,
when one explores their disc-averaged impact on galaxy scaling relations
and/or galaxy evolution (e.g., Obreschkow \& Glazebrook 2014; Obreschkow et
al.\ 2016; Zasov \& Zaitseva 2017; Kurapati et al.\ 2018; Romeo \& Mogotsi
2018; Romeo 2020; Romeo et al.\ 2020).  Our work, besides providing fresh
insights into popular scaling relations and discovering new ones, has imposed
tight constraints on the values of Toomre's (1964) $Q$ stability parameter
into which disc galaxies evolve (Romeo \& Mogotsi 2018; Romeo 2020; Romeo et
al.\ 2020).  The most basic result is that $Q$ is on average well above
unity, regardless of which disc component one considers: stars, atomic gas or
molecular gas (see fig.\ 1 of Romeo 2020).

In this paper, we explore $j$ and its connection with galaxy morphology, bar
structure and disc gravitational instability for disc galaxies of all
morphological types, from lenticulars to blue compact dwarfs, thus spanning
several orders of magnitude in stellar mass
($M_{\star}\approx10^{6.5\mbox{--}11.5}\,\mbox{M}_{\odot}$), atomic gas mass
($M_{\mathrm{HI}}\approx10^{7\mbox{--}11}\,\mbox{M}_{\odot}$), baryonic mass
($M_{\mathrm{b}}\approx10^{7.5\mbox{--}11.5}\,\mbox{M}_{\odot}$) and halo
mass ($M_{\mathrm{h}}\approx10^{9\mbox{--}13}\,\mbox{M}_{\odot}$).  The rest
of the paper is organized as follows.  In Sect.\ 2, we describe the galaxy
sample, data and statistics.  In Sect.\ 3, we explore the conservation of
specific angular momentum from a phenomenological point of view.  We analyse
this problem in detail by comparing not only stars and baryons (Sect.\ 3.1),
but also atomic gas and stars (Sect.\ 3.2), and discuss what physical
mechanisms are behind the observed correlations (Sect.\ 3.3).  In Sect.\ 4,
we explore whether barred galaxies are characterized by values of $j$ that
are systematically different from those of non-barred galaxies, as predicted
for instance by popular bar instability criteria.  We discuss this issue not
only in the context of bar instability (Sect.\ 4.1), but also in the context
of another important galaxy evolution process: the self-regulation of galaxy
discs driven by local gravitational instabilities (Sect.\ 4.2).  Finally, in
Sect.\ 5, we draw the conclusions of our work and point out their importance
for semi-analytic modelling of galaxy formation and evolution.

\section{METHODS}

\subsection{Galaxy sample}

We use a sample of 91 disc galaxies that stretch across all morphological
types, from lenticulars to blue compact dwarfs, and span a range of five
orders of magnitude in stellar mass
($M_{\star}\approx10^{6.5\mbox{--}11.5}\,\mbox{M}_{\odot}$) and four orders
of magnitude in atomic gas mass
($M_{\mathrm{HI}}\approx10^{7\mbox{--}11}\,\mbox{M}_{\odot}$), baryonic mass
($M_{\mathrm{b}}\approx10^{7.5\mbox{--}11.5}\,\mbox{M}_{\odot}$) and halo
mass ($M_{\mathrm{h}}\approx10^{9\mbox{--}13}\,\mbox{M}_{\odot}$).  Our
sample contains 77 galaxies of type S0--BCD from the `\emph{Spitzer}
Photometry and Accurate Rotation Curves' sample (SPARC; Lelli et al.\ 2016),
and 14 galaxies of type Im from the `Local Irregulars That Trace Luminosity
Extremes, The H\,\textsc{i} Nearby Galaxy Survey' (LITTLE THINGS; Hunter et
al.\ 2012).  Like the two parent samples, our galaxy sample is neither
statistically complete nor volume-limited, but it is nevertheless
representative of the full population of (regularly rotating) nearby
late-type galaxies (SPARC), with an emphasis on the faint end of the
luminosity function (LITTLE THINGS).

As a data set, our galaxy sample is the intersection of the samples analysed
by Romeo et al.\ (2020) and Mancera Pi\~{n}a et al.\ (2021a).  As such, it is
one of the largest samples of galaxies with reliable and quality-assessed
measurements of the following quantities, which are of key importance for our
analysis: the halo mass, $M_{\mathrm{h}}$, the stellar mass, $M_{\star}$, the
stellar specific angular momentum, $j_{\star}\equiv J_{\star}/M_{\star}$, the
mass of atomic hydrogen + helium gas, $M_{\mathrm{HI}}$, and the specific
angular momentum of atomic hydrogen + helium gas, $j_{\mathrm{HI}}\equiv
J_{\mathrm{HI}}/M_{\mathrm{HI}}$.

\subsection{Data}

$M_{\mathrm{h}}$ and $M_{\star}$ are taken from Posti et al.\ (2019) for
SPARC galaxies, and from Read et al.\ (2017) for LITTLE THINGS galaxies.  In
both cases, $M_{\mathrm{h}}$ and $M_{\star}$ were measured via rotation curve
decomposition, albeit adopting different halo models and different Bayesian
approaches to fit the observed rotation curves.  Such $M_{\mathrm{h}}$
measurements compare well with other recent determinations of
$M_{\mathrm{h}}$ made using different rotation curve decomposition methods,
both in the case of SPARC galaxies (e.g., Li et al.\ 2020) and in the case of
LITTLE THINGS galaxies (e.g., Mancera Pi\~{n}a et al.\ 2022).

$j_{\star}$ is taken from Posti et al.\ (2018) for SPARC galaxies, and from
Romeo et al.\ (2020) for LITTLE THINGS galaxies.  In the case of SPARC
galaxies, $j_{\star}$ was measured via radial integration, imposing a
convergence criterion on the cumulative $j_{\star}(<R)$ profile and including
asymmetric drift corrections (see, e.g., Binney \& Tremaine 2008,
chap.\ 4.8.2).  In the case of LITTLE THINGS galaxies, $j_{\star}$ was
measured adopting a commonly used approximation,
$j_{\star}=2\,R_{\mathrm{d}}V_{\mathrm{flat}}$, where $R_{\mathrm{d}}$ is the
exponential disc scale length and $V_{\mathrm{flat}}$ is the velocity along
the flat part of the rotation curve (e.g., Romanowsky \& Fall 2012).
$R_{\mathrm{d}}$ and $V_{\mathrm{flat}}$ were taken from Hunter \& Elmegreen
(2006) and Iorio et al.\ (2017), respectively.  In either case, $j_{\star}$
does not explicitly take into account the specific angular momentum of bars
or other non-axisymmetric structures; it only takes into account their
azimuthally averaged effect, consistent with all the gravitational
instability diagnostics used in this paper.

$M_{\mathrm{HI}}$ and $j_{\mathrm{HI}}$ are taken from Mancera Pi\~{n}a et
al.\ (2021a) for both SPARC and LITTLE THINGS galaxies.  $M_{\mathrm{HI}}$
was measured by integrating each H\,\textsc{i} surface density profile out to
the last observed radius, and by including the contribution of helium to the
atomic gas mass through a correction factor (1.33).  $j_{\mathrm{HI}}$ was
also measured via radial integration, imposing a convergence criterion on the
cumulative $j_{\mathrm{HI}}(<R)$ profile.  By construction, $j_{\mathrm{HI}}$
(like $j_{\star}$) includes the contribution of bars and other
non-axisymmetric structures only in an azimuthally averaged sense, consistent
with the assumptions behind the analysis carried out in this paper.

Note that all such measurements are based on high-quality rotation curves
that were derived from the same type of data (H\,\textsc{i} interferometric
observations) using consistent techniques (tilted ring models).  Therefore we
do not expect any significant bias.  Using such measurements, we compute the
baryonic mass, $M_{\mathrm{b}}$, and the baryonic specific angular momentum,
$j_{\mathrm{b}}\equiv J_{\mathrm{b}}/M_{\mathrm{b}}$, as
\begin{equation}
M_{\mathrm{b}}=
M_{\star}+M_{\mathrm{HI}}\,,
\end{equation}
\begin{equation}
j_{\mathrm{b}}=
\frac{j_{\star}M_{\star}+j_{\mathrm{HI}}M_{\mathrm{HI}}}
     {M_{\star}+M_{\mathrm{HI}}}\,.
\end{equation}
We neglect the contribution of molecular gas because it is relatively small
(e.g., Mancera Pi\~{n}a et al.\ 2021b), and because CO data are not available
for most galaxies of our sample (e.g., Hunter et al.\ 2012; Lelli et
al.\ 2016).

Another quantity that is of key importance for our analysis is the halo
specific angular momentum, $j_{\mathrm{h}}\equiv
J_{\mathrm{h}}/M_{\mathrm{h}}$.  Since this is not a truly observable galaxy
property, it is common practice to `measure' $j_{\mathrm{h}}$ via the
relation $j_{\mathrm{h}}\propto\lambda\,M_{\mathrm{h}}^{2/3}$, where
$\lambda$ is the halo spin parameter (e.g., Romanowsky \& Fall 2012;
Obreschkow \& Glazebrook 2014; Lapi et al.\ 2018; Okamura et al.\ 2018).
This is motivated by the fact that, in contrast to $j_{\mathrm{h}}$ itself,
$\lambda$ has been tightly constrained by $\Lambda$CDM simulations.  In fact,
$\lambda$ is well characterized by a log-normal probability distribution,
\begin{equation}
p(\lambda)\,\mbox{d}\lambda=
\frac{1}{\sqrt{2\pi}\sigma}
\exp\left[-\frac{(\ln\lambda-\ln\lambda_{0})^{2}}{2\sigma^{2}}\right]
\frac{\mbox{d}\lambda}{\lambda}\,,
\end{equation}
whose median $\lambda_{0}\approx0.035$ and width $\sigma\approx0.50$ (0.22
dex) do not depend significantly on halo mass, redshift, environment or
cosmology (e.g., Bullock et al.\ 2001; Macci\`{o} et al.\ 2007, 2008;
Rodr\'{i}guez-Puebla et al.\ 2016; Zjupa \& Springel 2017).  In view of this
fact, we too measure $j_{\mathrm{h}}$ via the
$j_{\mathrm{h}}\propto\lambda\,M_{\mathrm{h}}^{2/3}$ relation, which is fully
specified by Eq.\ (3) and the following equations:
\begin{equation}
j_{\mathrm{h}}=
\sqrt{2}\,\lambda\,R_{\mathrm{vir}}V_{\mathrm{vir}}\,,
\end{equation}
\begin{equation}
R_{\mathrm{vir}}=
\left(\frac{2}{\Delta_{\mathrm{c}}}
      \frac{GM_{\mathrm{h}}}{H_{0}^{2}}\right)^{1/3}\,,
\end{equation}
\begin{equation}
V_{\mathrm{vir}}=
\left(\frac{GM_{\mathrm{h}}}{R_{\mathrm{vir}}}\right)^{1/2}\,.
\end{equation}
Here $\lambda$ is the halo spin parameter redefined by Bullock et al.\ (2001)
that we have discussed above, $R_{\mathrm{vir}}$ and $V_{\mathrm{vir}}$ are
the halo virial radius and velocity (see, e.g., Cimatti et al.\ 2020),
$\Delta_{\mathrm{c}}$ is the critical overdensity for virialization, $H_{0}$
is the Hubble constant, and $G$ is the gravitational constant.  More
specifically, we set $H_{0}=67.4\;\mbox{km\,s}^{-1}\;\mbox{Mpc}^{-1}$ (Planck
Collaboration VI 2020) and $\Delta_{\mathrm{c}}=200$ in Eq.\ (5),
$\lambda_{0}=0.035$ and $\sigma=0.50$ (0.22 dex) in Eq.\ (3), and make use of
this equation to randomly generate one value of $\lambda$ for each galaxy of
our sample.  We then compute $j_{\mathrm{h}}$ from Eq.\ (4).

Our approach departs from the common practice of using the same value of
$\lambda$ for all the galaxies of the sample ($\lambda=\lambda_{0}$), so we
have tested it in Appendix B.  Our test demonstrates that varying the random
realization of $\lambda$ has a weak ($\la10\%$) effect on the results,
whereas suppressing the natural variance of $\lambda$ artificially constrains
the correlations between $j_{\mathrm{h}}$ and other fundamental galaxy
properties like $M_{\mathrm{h}}$ and $M_{\mathrm{b}}$ (see Fig.\ B1).

In addition to the key quantities specified above, we need to quantify the
morphological type of each galaxy, and to know whether a galaxy is barred or
non-barred.  The morphological type is taken from Lelli et al.\ (2016) for
SPARC galaxies and from Hunter et al.\ (2012) for LITTLE THINGS galaxies.
Information about the presence/absence of a bar is missing from the two
references above.  Therefore we extract it from HyperLeda (Makarov et
al.\ 2014), and classify the galaxies of our sample as `barred' (43\%) or
`non-barred' (47\%) on the basis of works referenced in that database, most
notably: (i) the `Third Reference Catalogue of Bright Galaxies' (RC3; de
Vaucouleurs et al.\ 1991), which is the primary frame of reference for
morphological classification of galaxies; and (ii) the `Galaxy Zoo 2' (GZ2;
Willett et al.\ 2013), which is a citizen science project with morphological
classifications of more than 300\,000 galaxies drawn from the Sloan Digital
Sky Survey (SDSS).  For some galaxies no consensus has been reached, so we
classify them as `uncertain' (10\%).  The fractions of barred and non-barred
galaxies that characterize our sample are consistent with those found by
G\'{e}ron et al.\ (2021) using the newest version of Galaxy Zoo, and with
their finding that there is a continuum of bar types, which varies from
`weakest' to `strongest'.%
\footnote{It is worth mentioning that all current methods of bar detection
  are subject to several observational effects difficult to quantify
  (bandpass, spatial resolution, imaging depth, etc.), and that there is an
  ongoing effort to evaluate and minimize such effects (e.g., Willett et
  al.\ 2013; Consolandi 2016; Abraham et al.\ 2018; G\'{e}ron et al.\ 2021).}

\subsection{Statistics}

To extract reliable information from our data, we use a variety of
statistical diagnostics, in particular several robust statistics.  These are
especially useful when the data are few or contain a significant fraction of
outliers, or even when the data deviate significantly from a normal
distribution (see, e.g., Rousseeuw 1991; Press et al.\ 1992, chap.\ 15.7).
Two eloquent examples of robust statistics are the median and the median
absolute deviation (MAD), which provide reliable estimates of the `central
value' and the `width' of a data set even when almost 50\% of the data are
outliers, contrary to the mean and the standard deviation.  Another example
of robust statistical methods is fitting a line to a set of data points by
minimizing their average absolute deviation from the line, a problem that is
solved by computing the median of the deviations (see pp.\ 698--700 of Press
et al.\ 1992).  If the data contain outliers, which is almost always the
case, then such `robust median-based' fitting provides more reliable results
than linear least-squares fitting (see fig.\ 15.7.1 of Press et al.\ 1992,
and figs 3--7 of Rousseeuw 1991).  Robust statistics are used not only in the
statistical description and modelling of scientific data (see, e.g.,
Feigelson \& Babu 2012) but also in data processing, where they are an
integral part of widespread techniques like Kalman filtering, median
filtering and wavelet-based denoising (see, e.g., Romeo 2021).

In this paper, we model the data using robust median-based fits (subroutine
\textsc{medfit.f} from Press et al.\ 1992), and measure the dispersion of the
data points around the model using a robust estimator of the $1\sigma$
scatter:
\begin{equation}
\mbox{SD}_{\mathrm{rob}}=
\frac{1}{0.6745}\times\mbox{MAD}\,,
\end{equation}
where $\mbox{SD}_{\mathrm{rob}}$ is the robust counterpart of the standard
deviation (see, e.g., M\"{u}ller 2000).  Values of $\mbox{SD}_{\mathrm{rob}}$
that are much less than the dynamic range of the data mean a tight relation.
When it is needed, we decompose the robust standard deviation from the model
into `bias' (median offset from the model) and `variance' (robust standard
deviation from the median trend), and estimate the uncertainty in the median
as follows:
\begin{equation}
\mbox{SE}_{\mathrm{rob}}=
1.253\times\frac{\mbox{SD}_{\mathrm{rob}}}{\sqrt{N}}\,,
\end{equation}
where $\mbox{SE}_{\mathrm{rob}}$ is the robust counterpart of the standard
error and $N$ is the number of data points (see again M\"{u}ller 2000).

When median-based descriptors are not available, we present the results of
several statistical measures and associated tests.  In particular, we measure
the correlation strength and significance of galaxy properties using
Pearson's $r$, Spearman's $\rho$ and Kendall's $\tau$ correlation
coefficients, together with their significance levels $p_{r}$, $p_{\rho}$ and
$p_{\tau}$ (subroutines \textsc{pearsn.f}, \textsc{spear.f} and
\textsc{kendl1.f} from Press et al.\ 1992).  Values of
$r,\rho,\tau\approx(-)1$ and $p_{r},p_{\rho},p_{\tau}\approx0$ mean a strong
and significant (anti)correlation.

Note that the statistics described above do not take measurement
uncertainties into account.  This has nothing to do with the code used to
compute such statistics.  It is a characteristic of all sample statistics,
robust or not (see, e.g., Feigelson \& Babu 2012).  Note also that while
robust median-based fitting provides reliable estimates of the slope and the
intercept of the best-fitting line, $y=a+bx$, it does not provide their
uncertainties.  We supplement such information with reliable results from
linear least-squares fitting, although this is not as easy as it may seem.
Our first attempt was to use the popular subroutine \textsc{fitexy.f} from
Press et al.\ (1992), which takes measurement uncertainties into account and
returns uncertainties in $a$ and $b$.  Using the data described in Sect.\ 2.2
as input, \textsc{fitexy.f} produces fits with output values of
$\chi^{2}\gg\nu$, the number of degrees of freedom ($\nu=89$), and output
values of the goodness-of-fit probability $q\ll10^{-3}$.  Such values mean
that those fits are poor and, in particular, that the uncertainties in $a$
and $b$ are unreliable (see chaps 15.1--15.3 of Press et al.\ 1992).  In view
of that, we decided to use another subroutine from Press et al.\ (1992),
\textsc{fit.f} with $\texttt{mwt}=0$ on input, which redefines the
uncertainties in $x$ and $y$ so that $q=1$ and returns reliable uncertainties
in $a$ and $b$.

In Sects 3 and 4, we will provide all such statistical information mainly in
summary form and simplified notation.  In particular, we will report the
correlation strength of any two galaxy properties as $\alpha\mbox{--}\beta$,
where $\alpha$ and $\beta$ are the values of the smallest and largest
correlation coefficients ($r$, $\rho$ and $\tau$).  In addition, we will
report their correlation significance as $10^{-\gamma}$, where $-\gamma$ is
the order of magnitude of the significance levels ($p_{r}$, $p_{\rho}$ and
$p_{\tau}$).

\section{CONSERVED, OR NOT CONSERVED, THAT IS THE QUESTION}

As discussed in Sect.\ 1, $j$ is one of the most fundamental galaxy
properties.  Furthermore, $j_{\star}/j_{\mathrm{h}}$ and
$j_{\mathrm{b}}/j_{\mathrm{h}}$ are of great theoretical importance because
they measure the fractions of specific angular momentum retained by stars and
baryons, i.e.\ how well specific angular momentum is conserved in the process
of galaxy formation and evolution.  The galaxy sample described in Sect.\ 2
is especially appropriate for exploring this problem, thanks to the high
quality and wide dynamic range of the data.  In this section, we explore the
conservation of specific angular momentum from such a phenomenological point
of view, and highlight the novelty of our results.  In particular, we show
that there are important differences between stars and baryons (Sect.\ 3.1),
and even more between atomic gas and stars (Sect.\ 3.2).  Finally, we discuss
what physical mechanisms are behind the observed correlations (Sect.\ 3.3).

\begin{figure*}
\includegraphics[scale=1.13]{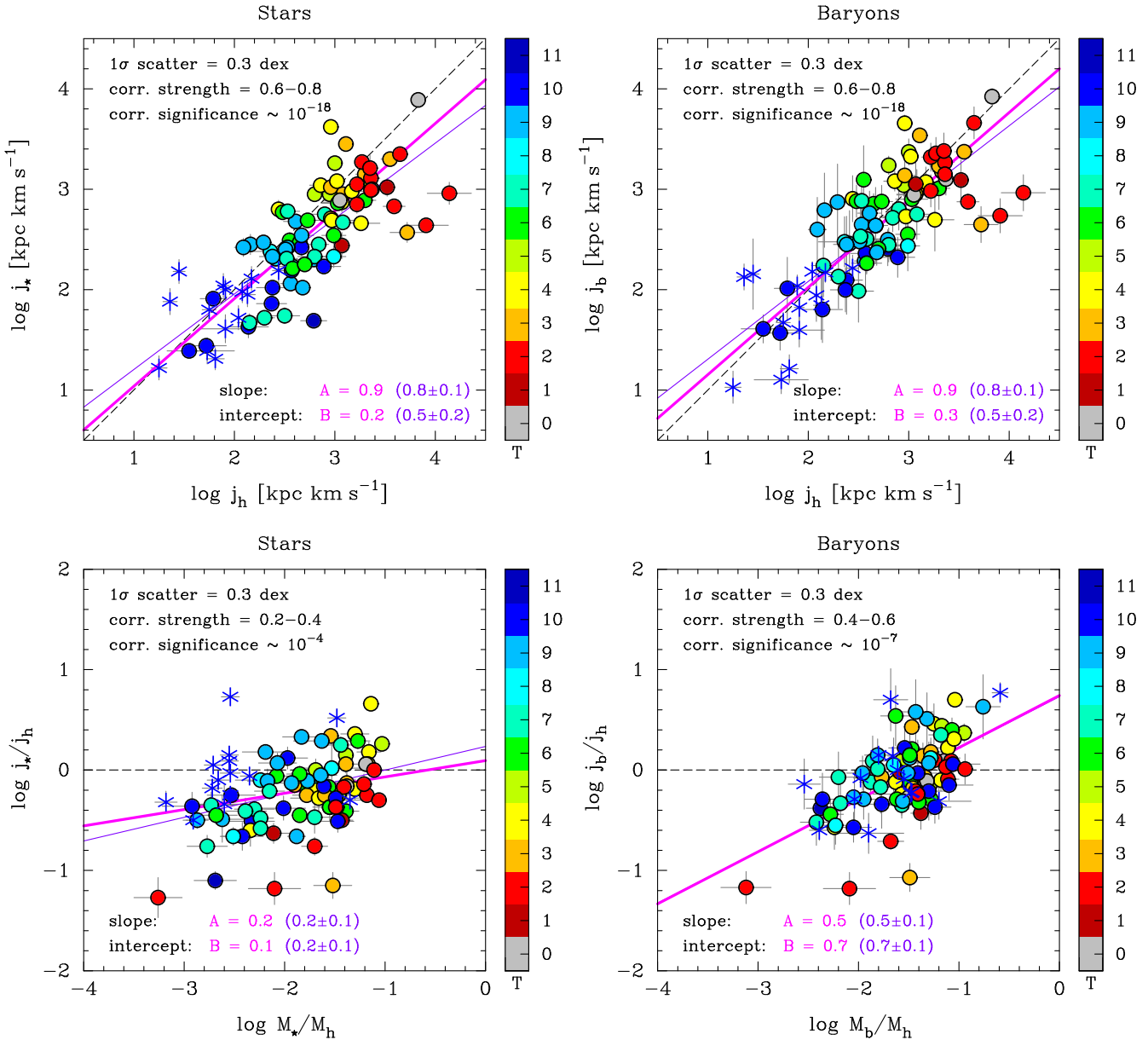}
\caption{Top panels: basic scaling relations between the stellar
  ($j_{\star}$), baryonic ($j_{\mathrm{b}}$) and halo ($j_{\mathrm{h}}$)
  specific angular momenta ($j\equiv J/M$) of disc galaxies.  Bottom panels:
  the strongest and most significant correlations between the retained
  fractions of specific angular momentum ($j_{\star}/j_{\mathrm{h}}$,
  $j_{\mathrm{b}}/j_{\mathrm{h}}$) and other galaxy properties.  Weaker and
  less significant correlations are shown in Figs A2 and A3.  The galaxy
  sample and the data are described in Sect.\ 2.  Galaxies are colour-coded
  by Hubble stage, and symbol-coded by their parent samples: SPARC (solid
  circles with black ouline) and LITTLE THINGS (asterisks).  The thick solid
  lines are robust median-based fits to the data points, while the thin solid
  lines are least-squares fits (see Sect.\ 2.3 for more information).  The
  dashed lines indicate conservation of specific angular momentum, i.e.\ that
  stars/baryons have retained the same amount of specific angular momentum as
  the host dark matter halo.  Statistical information about the data is given
  in summary form and simplified notation (see Sect.\ 2.3 for more
  information).}
\end{figure*}

\begin{figure*}
\includegraphics[scale=1.13]{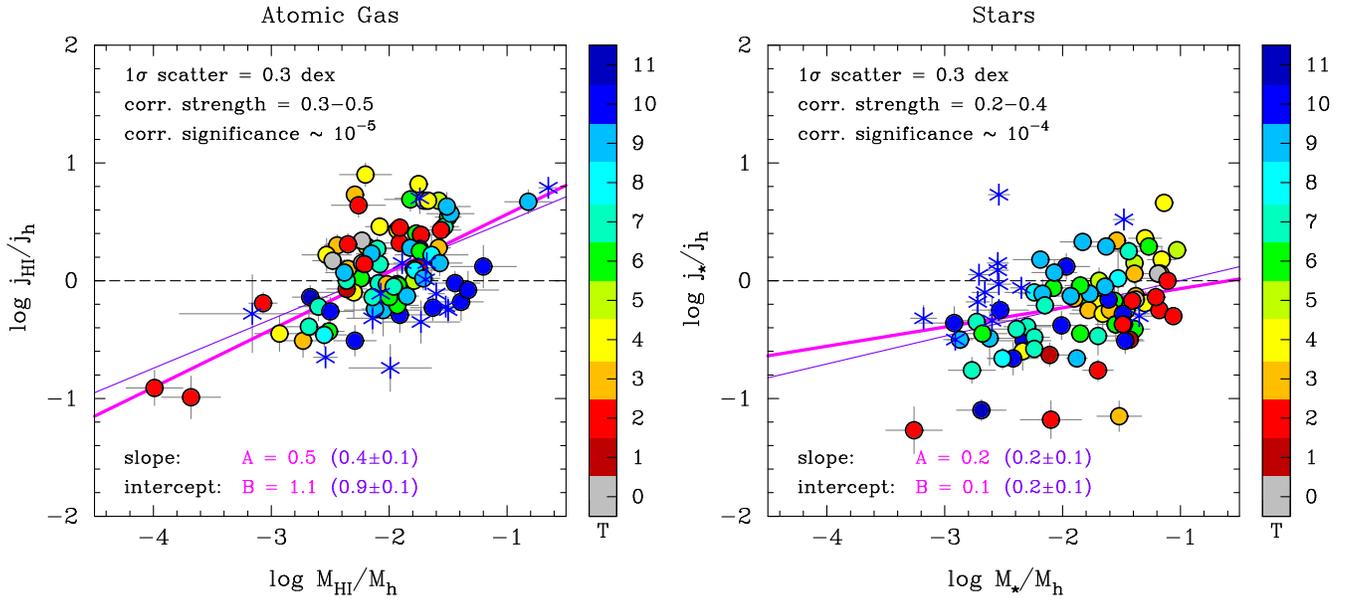}
\caption{The relation between retained fraction of specific angular momentum
  and mass fraction: atomic gas versus stars.  The galaxy sample and the data
  are described in Sect.\ 2.  Galaxies are colour-coded by Hubble stage, and
  symbol-coded by their parent samples: SPARC (solid circles with black
  ouline) and LITTLE THINGS (asterisks).  The thick solid lines are robust
  median-based fits to the data points, while the thin solid lines are
  least-squares fits (see Sect.\ 2.3 for more information).  The dashed lines
  indicate conservation of specific angular momentum, i.e.\ that
  atomic-gas/stars have retained the same amount of specific angular momentum
  as the host dark matter halo.  Statistical information about the data is
  given in summary form and simplified notation (see Sect.\ 2.3 for more
  information).}
\end{figure*}

\subsection{Stars versus baryons}

Fig.\ 1 shows basic scaling relations between $j_{\star}$, $j_{\mathrm{b}}$
and $j_{\mathrm{h}}$ (top panels), as well as the strongest and most
significant correlations between $j_{\star}/j_{\mathrm{h}}$,
$j_{\mathrm{b}}/j_{\mathrm{h}}$ and other galaxy properties (bottom panels).
Each panel also shows a robust median-based fit to the data points (thick
solid line), a linear least-squares fit (thin solid line), the locus of
points where specific angular momentum is conserved (dashed line), as well as
the results of various statistical diagnostics (see Sect.\ 2.3 for more
information about the statistics and, in particular, about the fitting
methods).  The top panels of Fig.\ 1 illustrate that $j_{\star}$ and
$j_{\mathrm{b}}$ are tightly related to $j_{\mathrm{h}}$, and that the two
relations have similar logarithmic slopes, $A\approx1$, and intercepts,
$B\approx0$.  So specific angular momentum is approximately conserved in a
statistical sense.  But how good is this approximation?  And are the retained
fractions of specific angular momentum subject to systematic effects?  The
bottom panels of Fig.\ 1 provide explicit answers to these questions, as
pointed out below.
\begin{enumerate}
\item \textbf{How well is specific angular momentum conserved in a
  statistical sense?}  It depends on whether we consider the baryons or only
  the stars.  In fact, while in both cases the distribution of data points is
  offset towards negative logarithmic values, the median of
  $j_{\mathrm{b}}/j_{\mathrm{h}}$ ($0.81\pm0.07$) is well above the median of
  $j_{\star}/j_{\mathrm{h}}$ ($0.63\pm0.06$).  Hence, on average, specific
  angular momentum is conserved to better than 20\% for baryons and to within
  40\% for stars.  Note that such estimates are fully meaningful, regardless
  of how strongly or significantly the retained fractions of specific angular
  momentum correlate with other galaxy properties.  This is because the
  probability distributions of $\log\,j_{\mathrm{b}}/j_{\mathrm{h}}$ and
  $\log\,j_{\star}/j_{\mathrm{h}}$ are clearly unimodal and more peaked than
  a Gaussian (see Fig.\ A1 for detailed statistical information), and because
  the median is a robust estimator of the central value of a distribution if
  this has a strong central tendency (see chap.\ 14.1 of Press et al.\ 1992).
\item \textbf{Are the retained fractions of specific angular momentum subject
  to systematic effects?}  Yes, they are, especially the baryonic one.  There
  is in fact a moderately strong (e.g.\ $\rho\approx0.5$) but very
  significant ($p_{\rho}\sim10^{-7}$) correlation between
  $j_{\mathrm{b}}/j_{\mathrm{h}}$ and $M_{\mathrm{b}}/M_{\mathrm{h}}$.  The
  stellar counterpart of this correlation, $j_{\star}/j_{\mathrm{h}}$ vs
  $M_{\star}/M_{\mathrm{h}}$, is also significant ($p_{\rho}\sim10^{-4}$)
  although weaker ($\rho\approx0.4$).  Thus the retained fractions of
  specific angular momentum do depend systematically on the galaxy formation
  and star formation efficiencies, and vary on average as
  $j_{\mathrm{b}}/j_{\mathrm{h}}\propto
  (M_{\mathrm{b}}/M_{\mathrm{h}})^{0.5}$ and $j_{\star}/j_{\mathrm{h}}\propto
  (M_{\star}/M_{\mathrm{h}})^{0.2}$.  Both robust median-based fitting and
  linear least-squares fitting yield the same systematic dependence on mass
  fraction, at least within the parameter uncertainties.
\end{enumerate}

Previous investigations focused on either $j_{\star}/j_{\mathrm{h}}$ (e.g.,
Romanowsky \& Fall 2012; Lapi et al.\ 2018) or
$j_{\mathrm{b}}/j_{\mathrm{h}}$ (e.g., Dutton \& van den Bosch 2012; Mancera
Pi\~{n}a et al.\ 2021a), and found similar results:
$j_{\star}/j_{\mathrm{h}}\approx0.6\mbox{--}1.0$, nearly independent of
$M_{\mathrm{h}}$ or $M_{\star}$;
$j_{\mathrm{b}}/j_{\mathrm{h}}\approx0.5\mbox{--}0.7$, nearly independent of
$M_{\mathrm{h}}$ or $M_{\mathrm{b}}$.

Our comparative analysis has instead revealed noteworthy and previously
undetected differences between $j_{\star}/j_{\mathrm{h}}$ and
$j_{\mathrm{b}}/j_{\mathrm{h}}$.  Note, in particular, that it is highly
non-trivial to detect and differentiate the systematic effects pointed out
above (item ii).  It requires not only high-quality data with a wide dynamic
range, but also a detailed comparative analysis that stretches across a
variety of galaxy properties.  In fact, $j_{\mathrm{b}}/j_{\mathrm{h}}$ and
$j_{\star}/j_{\mathrm{h}}$ do not show any particularly significant
($p\la10^{-4}$) correlation with basic properties like $j_{\mathrm{h}}$,
$M_{\mathrm{h}}$ or their baryonic/stellar counterparts (see Figs A2 and A3).
Note also that the baryonic and stellar scaling relations pointed out above
(item ii) are basically consistent with `inside-out' or `biased-collapse'
models of galaxy formation (e.g., Dutton \& van den Bosch 2012; Romanowsky \&
Fall 2012).  In other words, those models assume that there is a power-law
relation between retained fraction of specific angular momentum and mass
fraction, but they do not predict the actual slopes of the baryonic and
stellar relations, which most likely result from the galaxy evolution
processes involved in the gas-star cycle.

\subsection{Atomic gas versus stars}

To further understand how well specific angular momentum is conserved in the
process of galaxy formation and evolution, let us finally turn our attention
to atomic gas and analyse $j_{\mathrm{HI}}/j_{\mathrm{h}}$.

Fig.\ 2 shows $j_{\mathrm{HI}}/j_{\mathrm{h}}$ vs
$M_{\mathrm{HI}}/M_{\mathrm{h}}$ face to face with its stellar counterpart.
The main results of our comparative analysis are pointed out below.
\begin{enumerate}
\item \textbf{Basic constraints.}  The most striking result is that atomic
  gas has actually \emph{gained} more specific angular momentum than the host
  dark matter halo.  The median of $j_{\mathrm{HI}}/j_{\mathrm{h}}$
  ($1.23\pm0.14$) is in fact well above unity, and indeed twice as large as
  the median of $j_{\star}/j_{\mathrm{h}}$.  Such estimates are meaningful
  because the probability distribution of
  $\log\,j_{\mathrm{HI}}/j_{\mathrm{h}}$ has a strong central tendency,
  although not as strong as the one shown by stars (see Fig.\ A1 for detailed
  statistical information).
\item \textbf{Systematic trends.}  Concerning the relation between `retained'
  fraction of specific angular momentum and mass fraction, atomic gas shows a
  steeper scaling than stars, $j_{\mathrm{HI}}/j_{\mathrm{h}}\propto
  (M_{\mathrm{HI}}/M_{\mathrm{h}})^{0.5}$, and a slightly higher degree of
  correlation (e.g.\ $\rho\approx0.4$ and $p_{\rho}\sim10^{-5}$).  Note also
  that the two relations show opposite residual trends with galaxy
  morphology.  For instance, early-type galaxies tend to cluster above
  (below) the best-fitting relation found for atomic gas (stars), hence they
  tend to have higher $j_{\mathrm{HI}}/j_{\mathrm{h}}$ (lower
  $j_{\star}/j_{\mathrm{h}}$) than predicted.  This tendency is reversed for
  late-type galaxies.  It is most likely because of such opposite residual
  trends that baryons show a higher degree of correlation than stars and
  atomic gas.  In fact, $j_{\mathrm{b}}$ is the mass-weighted average of
  $j_{\star}$ and $j_{\mathrm{HI}}$ (see Eq.\ 2), which tends to cancel out
  opposite trends.
\end{enumerate}

Our result (i) is consistent with two results from cosmological simulations
of galaxy formation, namely that accreting gas has higher specific angular
momentum than the dark matter halo (Kimm et al.\ 2011; Stewart et al.\ 2013),
and that gas in galaxy discs tends to have higher specific angular momentum
than stars (Teklu et al.\ 2015; Agertz \& Kravtsov 2016; El-Badry et
al.\ 2018).  Our finding that
$(j_{\mathrm{HI}}/j_{\mathrm{h}})\approx2\,(j_{\star}/j_{\mathrm{h}})$ is
also consistent with a result previously found by Mancera Pi\~{n}a et
al.\ (2021b), namely that $j_{\mathrm{HI}}/j_{\star}\approx2$.  Those authors
showed that $j_{\mathrm{HI}}/j_{\star}$ is in fact independent of
$M_{\mathrm{b}}$, and that its value regulates the slopes of the relations
between $j$, $M$ and gas fraction.

As regards the relations between retained fraction of specific angular
momentum and basic galaxy properties, atomic gas and stars show similarities
as well as significant differences (cf.\ Figs A4 and A3, respectively).  On
the one hand, $j_{\mathrm{HI}}/j_{\mathrm{h}}$ does not correlate with
$j_{\mathrm{h}}$ or $M_{\mathrm{h}}$, like $j_{\star}/j_{\mathrm{h}}$.  On
the other hand, $j_{\mathrm{HI}}/j_{\mathrm{h}}$ has a moderately high degree
of correlation with $j_{\mathrm{HI}}$ and $M_{\mathrm{HI}}$, while
$j_{\star}/j_{\mathrm{h}}$ is uncorrelated with $j_{\star}$ or $M_{\star}$.

\subsection{What physical mechanisms are behind the observed correlations?}

One of the most innovative results of our analysis is the finding that
galaxies with larger baryon fractions have also retained larger fractions of
their specific angular momentum (see Sects 3.1 and 3.2).  In this section, we
discuss what physical mechanisms are behind the observed correlations (bottom
panels of Fig.\ 1, and Fig.\ 2).

Indeed, such correlations impose important constraints on the physics
governing the galactic outflow-accretion cycle across galaxy masses.  Dutton
\& van den Bosch (2012) argued that the empirical scaling relations between
$j_{\mathrm{b}}/j_{\mathrm{h}}$, $M_{\mathrm{b}}/M_{\mathrm{h}}$ and
$M_{\mathrm{h}}$ require three ingredients: (i) galactic outflows, driven by
stellar and/or AGN feedback; (ii) angular momentum transfer from accreting
gas to the dark matter halo, driven by dynamical friction; and, most
importantly, (iii) that the efficiency of angular momentum loss decreases
with increasing halo mass.  State-of-the-art cosmological simulations of
galaxy formation are providing valuable insights into the angular momentum of
accreting gas.  Hafen et al.\ (2022) show that the circumgalactic medium
(CGM) inside massive haloes ($M_{\mathrm{h}}\sim10^{12}\,\mbox{M}_{\odot}$)
tends to be virialized, as the cooling time is longer than the dynamical
time.  This causes gas to accrete via hot rotating flows, which feed the
outer regions of disc galaxies by transferring angular momentum that is often
aligned with the angular momentum of the disc (see also Trapp et al.\ 2022).
Hafen et al.\ (2022) also show that the CGM inside lower-mass haloes
($M_{\mathrm{h}}\sim10^{11}\,\mbox{M}_{\odot}$) is less, or not at all,
virialized.  This causes the angular momentum of accreting gas to be often
misaligned with that of the disc.  Since accreting gas transfers angular
momentum to both the disc and the dark matter halo, the two findings above
suggest that the efficiency with which angular momentum is transferred to the
halo decreases with increasing halo mass, and so does the efficiency of
angular momentum loss.  The two findings above are therefore in qualitative
agreement with the theoretical arguments of Dutton \& van den Bosch (2012),
as well as with our observed correlations.

\section{BARRED VERSUS NON-BARRED GALAXIES}

In this section, we explore the connection between the specific angular
momentum of disc galaxies, bar structure and disc gravitational instability.
The core of the problem is whether barred galaxies are characterized by
values of $j$ that are systematically different from those of non-barred
galaxies, as predicted for instance by popular bar instability criteria.
Below we discuss this issue not only in the context of bar instability (Sects
4.1.1--4.1.3), but also in the context of another important galaxy evolution
process: the self-regulation of galaxy discs driven by local gravitational
instabilities (Sects 4.2.1--4.2.2).

\subsection{Observational test of the Efstathiou, Lake \& Negroponte (1982)
            bar instability criterion}

\subsubsection{Overview}

A decade after the pioneering work of Ostriker \& Peebles (1973), Efstathiou
et al.\ (1982) formulated a simple bar instability criterion in terms of
observable galaxy properties (hereafter ELN criterion):
\begin{equation}
\mathcal{E}\equiv
\frac{V_{\mathrm{max}}}
     {(GM_{\mathrm{d}}/R_{\mathrm{d}})^{1/2}}
\;\la\;1\,,
\end{equation}
where $V_{\mathrm{max}}$ is the maximum rotation velocity, $M_{\mathrm{d}}$
is the mass of the disc, and $R_{\mathrm{d}}$ is the exponential disc scale
length.  The instability threshold is $\simeq1.1$ for stellar discs
(Efstathiou et al.\ 1982) and $\simeq0.9$ for gas discs (Christodoulou et
al.\ 1995), but it is common to approximate these thresholds more simply as
$\approx1$.  Mo et al.\ (1998) and van den Bosch (1998) did so, and used a
detailed disc formation model to reformulate the ELN criterion in terms of
more fundamental galaxy properties: the disc mass fraction,
$M_{\mathrm{d}}/M_{\mathrm{h}}$, and the disc spin parameter,
$\lambda\,(j_{\mathrm{d}}/j_{\mathrm{h}})$, i.e.\ the halo spin parameter
($\lambda$) times the fraction of specific angular momentum retained by the
disc ($j_{\mathrm{d}}/j_{\mathrm{h}}$).  The resulting bar instability
condition is more elaborate than Eq.\ (9), but Mo et al.\ (1998) showed that
such a condition depends weakly on the disc-halo model and is well
approximated by a simple formula:
\begin{equation}
\mathcal{E}^{2}\approx
\lambda\,
\frac{(j_{\mathrm{d}}/j_{\mathrm{h}})}
     {(M_{\mathrm{d}}/M_{\mathrm{h}})}
\;\la\;1\,,
\end{equation}
here expressed in explicit form using our notation.%
\footnote{Deriving Eq.\ (10) from Eq.\ (9) is a complex procedure, which
  involves several steps of the disc-halo modelling and several
  approximations of the model parameters.  The interested reader is referred
  to sects 2.2, 2.3 and 3.2 of Mo et al.\ (1998) for detailed information.}
Such a criterion predicts that a disc galaxy is bar unstable if and only if
the disc spin parameter is lower than the disc mass fraction.  This means
that if one disregards the galaxy evolution processes that follow the
formation of a bar, as is commonly done when comparing the predictions of bar
instability criteria with observations, then barred galaxies should all be
gravitationally unstable and characterized by values of
$j_{\mathrm{d}}/j_{\mathrm{h}}$ that are systematically lower than those of
non-barred galaxies (for a given $\lambda$ and a given
$M_{\mathrm{d}}/M_{\mathrm{h}}$).

Athanassoula (2008) pointed out two major limitations of the ELN criterion,
and illustrated them with eloquent simulation tests.  First of all, the ELN
criterion is based on 2D simulations so it does not take into account the
interaction between disc and halo, which has a strong destabilizing impact.
Secondly, the ELN criterion does not properly take into account the disc
velocity dispersion or the central concentration of the halo, either of which
has a stabilizing effect.  Indeed, the disc velocity dispersion, $\sigma$, is
one of the quantities that most radically affect the onset of gravitational
instabilities in galaxy discs, and the quantity that was most drastically
modelled in early (2D) simulations.  This concerns not only $\sigma_{z}$,
which gives vertical structure to the disc and plays an important stabilizing
role (Vandervoort 1970; Romeo 1992, 1994), but also $\sigma_{R}$, whose
stabilizing role can be critically impacted by low force resolution (Romeo
1994, 1997, 1998a,b).  Note also that $\sigma_{z}/\sigma_{R}$ is an important
parameter for the evolution of a bar: values of $\sigma_{z}/\sigma_{R}\la0.3$
cause the bending instability (buckling of the bar), which also causes the
formation of boxy/peanut structures (see Rodionov \& Sotnikova 2013 for a
recent overview and detailed analysis).  All that is not (properly) taken
into account by the ELN criterion.

Athanassoula (2008) also mentioned another limitation of the ELN criterion,
namely that it does not take into account the multi-component nature of
galaxy discs.  In other words, the fact that Eqs (9) and (10) are valid for
discs made of \emph{either} stars \emph{or} gas does not mean that they can
be applied to discs made of \emph{both} stars \emph{and} gas, as is commonly
done.%
\footnote{This is one of the lessons learned in the context of local disc
  gravitational instabilities.  Look for instance at fig.\ 5 of Romeo \&
  Wiegert (2011), and see how dramatically the gas $Q$ parameter
  misrepresents the actual stability level of nearby star-forming spirals.}
In fact, zoom-in cosmological simulations show that high gas fractions tend
to dissolve bars (Kraljic et al.\ 2012).

Sellwood (2016) carried out further simulation tests that illustrated, once
again, the importance of disc-halo interaction for bar instability, thus the
inadequacy of the ELN criterion (see also Berrier \& Sellwood 2016).

In spite of such criticisms, the ELN criterion is used by all current
semi-analytic models of galaxy formation and evolution to `create' bulges in
disc galaxies that are predicted to be bar unstable (see sect.\ 1 of Devergne
et al.\ 2020 for an overview).  Indeed, the popularity of the ELN criterion
originates not only from its simplicity, but also from the belief that its
inaccuracy is `likely to be' negligible in comparison with other
uncertainties of the modelling, for example the mass of the bulge formed by
bar instability (see again sect.\ 1 of Devergne et al.\ 2020).

\begin{figure*}
\includegraphics[scale=1.25]{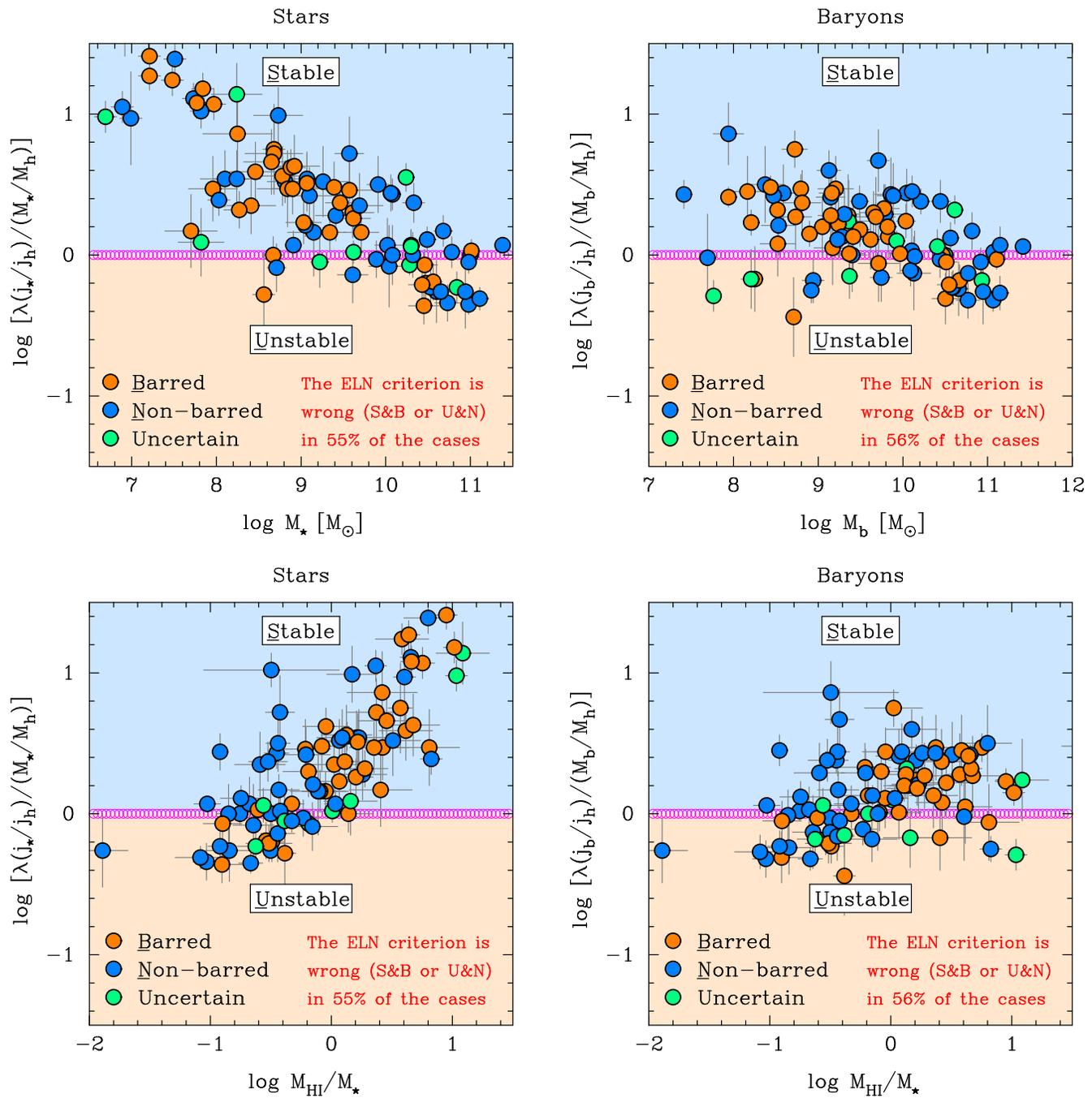}
\caption{Observational test of the Efstathiou, Lake \& Negroponte (1982) bar
  instability criterion (hereafter ELN criterion), as reformulated by Mo et
  al.\ (1998).  Such a criterion predicts that a disc galaxy is bar unstable
  if and only if
  $\lambda\,(j_{\mathrm{d}}/j_{\mathrm{h}})/(M_{\mathrm{d}}/M_{\mathrm{h}})
  \la1$, and is used by all current semi-analytic models of galaxy formation
  and evolution to `create' bulges in disc galaxies that are predicted to be
  bar unstable (see sect.\ 1 of Devergne et al.\ 2020 for an overview).  The
  galaxy sample and the data are described in Sect.\ 2.  The left and right
  panels illustrate our test for two popular implementations of the ELN
  criterion: one focusing on the stellar disc, and the other including the
  whole baryonic disc.  In each case, the ELN parameter is shown as a
  function of mass (top) and gas mass fraction (bottom).  If the ELN
  criterion was reliable, then the light orange/azure region would be almost
  entirely populated by the orange/azure data points.  Clearly, this is not
  the case in any of the panels.  Furthermore, there are clear systematic
  trends with stellar/baryonic mass and gas mass fraction (see Sect.\ 4.1.2
  for more information).}
\end{figure*}

\subsubsection{How accurate is the ELN criterion from a statistical point of
               view?}

This is a crucial question that naturally arises from the discussion above.
To answer this question, we test the ELN criterion observationally making use
of Eq.\ (10).  To the best of our knowledge, this is the first observational
test performed on the ELN criterion; and it is statistically unbiased, given
that the fractions of barred and non-barred galaxies that characterize our
sample are consistent with those found by G\'{e}ron et al.\ (2021) using the
newest version of Galaxy Zoo (see Sect.\ 2.2 for more information).  We make
use of Eq.\ (10), rather than Eq.\ (9), because it naturally connects with
the analysis carried out in Sect.\ 3.  To perform such a test, we disregard
the galaxy evolution processes that follow the formation of a bar, as is
commonly done when comparing the predictions of bar instability criteria with
observations (e.g., Efstathiou et al.\ 1982; Mo et al.\ 1998; van den Bosch
1998; Okamura et al.\ 2018; Kataria et al.\ 2020).

Fig.\ 3 illustrates our test for two popular implementations of the ELN
criterion: one focusing on the stellar disc (left panels), and the other
including the whole baryonic disc (right panels).%
\footnote{Remember from Sect.\ 2.2 that CO data are not available for most
  galaxies of our sample, so what we call `baryons' are stars and atomic gas.
  To neglect molecular gas may seem an oversimplification because bars can
  host large amounts of such gas (e.g., Renaud et al.\ 2015).  But remember
  also that the ELN criterion is a global bar instability condition, which
  concerns the disc as a whole, and that the contribution of molecular gas to
  $M_{\mathrm{d}}$ and $j_{\mathrm{d}}$ is relatively small (e.g., Mancera
  Pi\~{n}a et al.\ 2021b).}
In each case, the ELN parameter is shown as a function of mass (top) and gas
mass fraction (bottom).  The main results of our comparative analysis are
pointed out below.
\begin{enumerate}
\item \textbf{Overall accuracy.}  If the ELN criterion was reliable, then the
  light orange/azure region would be almost entirely populated by the
  orange/azure data points.  Clearly, this is not the case in any of the
  panels.  Indeed, barred and non-barred galaxies are mixed across the entire
  range of values spanned by
  $\lambda\,(j_{\mathrm{d}}/j_{\mathrm{h}})/(M_{\mathrm{d}}/M_{\mathrm{h}})$.
  This is a fundamental limitation of the ELN criterion, which one cannot
  overcome by shifting the instability threshold up or down.  To measure how
  inaccurate the ELN criterion is, we count how many galaxies fall within the
  `wrong' regime: `stable and barred', or `unstable and non-barred'.  This
  happens in about 55\% of the cases, regardless of the implementation.  In
  simple words, the ELN criterion has a fifty-fifty chance of being right or
  wrong.  This is not an artefact of using the ELN criterion reformulated by
  Mo et al.\ (1998), rather than the original ELN criterion itself: Eqs (9)
  and (10) are almost indistinguishable from a statistical point of view (see
  Fig.\ B2).  A further additional test shows that applying the ELN criterion
  to the atomic gas disc does not improve its accuracy with respect to the
  popular stellar-disc implementation (see Fig.\ B3).  Thus the overall
  inaccuracy of the ELN criterion is a robust result demonstrated by a
  detailed comparative analysis.
\item \textbf{Systematic trends.}  When the ELN criterion is applied to the
  stellar disc (see the left panels of Fig.\ 3), there are clear systematic
  trends with stellar mass and gas mass fraction across the entire ranges of
  values spanned by $M_{\star}$ and $M_{\mathrm{HI}}/M_{\star}$.  As a result
  of such trends, the ELN parameter is typically above unity for
  $M_{\star}\la10^{10}\,\mbox{M}_{\odot}$ or
  $M_{\mathrm{HI}}/M_{\star}\ga0.3$, and can be as high as 10 or more for
  $M_{\star}\la3\times10^{8}\,\mbox{M}_{\odot}$ or
  $M_{\mathrm{HI}}/M_{\star}\ga1$.  When instead the ELN criterion is applied
  to the baryonic disc (see now the right panels of Fig.\ 3), the trends are
  more moderate than in the previous case, but the ELN parameter is still
  typically above unity for $M_{\mathrm{b}}\la10^{10}\,\mbox{M}_{\odot}$ or
  $M_{\mathrm{HI}}/M_{\star}\ga0.3$.  Thus the ELN criterion tends to
  overpredict how bar stable disc galaxies are, especially in the case of
  low-mass gas-rich spirals and dwarfs.  Such a tendency is exacerbated when
  one uses the popular stellar-disc implementation of the ELN criterion.
\end{enumerate}

Our result (i) seems to be at odds with a recent result found by
Izquierdo-Villalba et al.\ (2022) using the IllustrisTNG simulations, namely
that the ELN criterion successfully identifies $\approx75\%$ of the strongly
barred galaxies and $\approx80\%$ of the non-barred ones.  Note, however,
that their result concerns Milky Way--type galaxies of stellar mass
$M_{\star}\approx10^{10.4\mbox{--}11.0}\,\mbox{M}_{\odot}$, and that the ELN
criterion is less inaccurate for $M_{\star}\ga10^{10}\,\mbox{M}_{\odot}$ than
at lower stellar masses (see the top-left panel of Fig.\ 3, and item ii).
The inaccuracy of the ELN criterion at low stellar masses was already
suspected by Irodotou et al.\ (2019) and Izquierdo-Villalba et al.\ (2019) in
the context of semi-analytic modelling (`L-Galaxies' model), and it was one
of the motivations for introducing improved versions of the ELN criterion
(Irodotou et al.\ 2019).

Our result (ii) not only demonstrates the inaccuracy of the ELN criterion at
low $M_{\star}$ ($M_{\mathrm{b}}$) or high $M_{\mathrm{HI}}/M_{\star}$ and
shows that it is caused by the systematic trends of the ELN parameter, but
also suggests how one could improve the accuracy of the ELN criterion: by
fitting such trends and subtracting the best-fitting relations from the ELN
parameter.  We will not do that because there would still be residual mixing
of barred and non-barred galaxies, as is clear from Fig.\ 3, and because we
do not believe that the complex phenomenology of bars in disc galaxies can be
encapsulated into a simple analytical criterion.

\subsubsection{A final comment on the complexity of the problem}

Now that we have highlighted the strength of our results, let us finally
remember their weakness.  As mentioned in Sect.\ 4.1.2, we have disregarded
the galaxy evolution processes that follow the formation of a bar, as is
commonly done when comparing the predictions of bar instability criteria with
observations (e.g., Efstathiou et al.\ 1982; Mo et al.\ 1998; van den Bosch
1998; Okamura et al.\ 2018; Kataria et al.\ 2020).  This could lead to
incorrect results.  For example, a galaxy could be unstable to bar formation
according to the ELN criterion but become stable after the bar has formed, as
a result of complex galaxy evolution processes such as the redistribution of
mass driven by the transfer of angular momentum from the bar to the outer
disc and the halo (see, e.g., Gadotti 2009; Combes 2011; Athanassoula 2013;
Kormendy 2013).  The influence of such processes on bar instability is a
highly non-trivial aspect of the problem, whose solution will demand
painstaking comparative analyses of observed and simulated galaxies,
considering that there is still tension between observations and simulations
as regards the evolution of basic bar properties (e.g., Kim et al.\ 2021;
Roshan et al.\ 2021; Lee et al.\ 2022).

\subsection{Self-regulation of galaxy discs driven by local gravitational
            instabilities}

\subsubsection{Overview}

Pioneering simulation work on spiral structure in galaxies predicted that
galaxy discs self-regulate to stability levels that are not far from the
critical threshold predicted by Toomre (1964),
$Q_{\mathrm{T}}\equiv\kappa\sigma_{R\star}/(3.36\,G\Sigma_{\star})\sim1$, and
that such a process is driven by local gravitational instabilities, gas
dissipation and other sources of dynamical heating/cooling (e.g., Miller et
al.\ 1970; Hohl 1971; Sellwood \& Carlberg 1984; Carlberg \& Sellwood 1985).
Today, several decades after such work, the self-regulation of galaxy discs
is still a hot topic.  On the one hand, there have been significant advances
in our understanding of the complex interplay between the heating and cooling
processes that lead to self-regulation (e.g., Bertin \& Romeo 1988; Romeo
1990; Cacciato et al.\ 2012; Forbes et al.\ 2012, 2014; Goldbaum et
al.\ 2015; Krumholz et al.\ 2018).  On the other hand, there is not yet a
broad understanding of how self-regulated galaxy discs are.  For instance,
several star formation models postulate the existence of a self-regulation
process that keeps gas close to marginal stability, i.e.\ they assume that
$Q_{\mathrm{gas}}\equiv\kappa\sigma_{\mathrm{gas}}/(\pi
G\Sigma_{\mathrm{gas}})\simeq1$ (see sect.\ 1 of Krumholz et al.\ 2018 for an
overview).  This is in sharp contrast to the observed radial distribution of
$Q_{\mathrm{gas}}$ in galaxy discs, which is remarkably unconstrained (Leroy
et al.\ 2008; Romeo \& Wiegert 2011).

To assess how self-regulated galaxy discs are, one must take into account
their multi-component nature and their vertical structure.  This can be done,
easily and accurately, by making use of the Romeo \& Falstad (2013)
$\mathcal{Q}_{\mathrm{RF}}$ stability parameter.  Romeo \& Mogotsi (2017,
2018) did so and, in spite of using different galaxy samples and different
statistical methods, they found a similar result:
$\mathcal{Q}_{\mathrm{RF}}\approx2$, with a scatter of $\approx$ 0.2 dex.
This means that galaxy discs are well self-regulated.  Indeed, the radial
distribution of $\mathcal{Q}_{\mathrm{RF}}$ is remarkably flat up to
galactocentric distances as large as the optical radius, and its median value
($\approx$ 2) is consistent with the destabilizing effects of
non-axisymmetric perturbations and gas dissipation (see fig.\ 3 and sect.\ 3
of Romeo \& Mogotsi 2017).  Similar results have been found using
state-of-the-art simulations of disc galaxy evolution (Renaud et al.\ 2021;
Ejdetj\"{a}rn et al.\ 2022).

To further understand how self-regulated galaxy discs are, one should analyse
in detail the building blocks of $\mathcal{Q}_{\mathrm{RF}}$, i.e.\ the $Q$
parameters of stars, atomic gas and molecular gas:
$Q_{i}=\kappa\sigma_{i}/\pi G\Sigma_{i}$
$(i=\star,\,\mbox{H\,\textsc{i}},\,\mbox{H}_{2})$.  Romeo (2020) did so and
showed that the radial distribution of $Q_{i}$ changes dramatically not only
from stars to gas, but also between the atomic and molecular gas phases (see
his fig.\ 1).  He also analysed the mass-weighted average of $Q_{i}$ over the
disc, $\langle Q_{i}\rangle$, and found that the median of $\langle
Q_{i}\rangle$ over the galaxy sample is $\approx$ 2--3 for stars and $\sim$
10 for atomic/molecular gas, while the $1\sigma$ scatter is $\approx$ 0.2 dex
for all the components.  This means that, despite the diverse phenomenology
of $Q$, galaxy discs are so well self-regulated that each disc component has
its own characteristic value of $\langle Q\rangle$.  Indeed, this is true for
disc galaxies of all morphological types, from lenticulars to blue compact
dwarfs, at least if one considers their stellar and atomic gas components
(see fig.\ 2 of Romeo et al.\ 2020 and fig.\ 4 of Romeo 2020, respectively).

Finally, note that there is a relation between $\langle Q\rangle$ and the ELN
parameter: $\langle Q\rangle\propto
j\sigma/GM\propto\mathcal{E}^{2}\,\sigma/V$ (Romeo \& Mogotsi 2018).  Hence
$\langle Q\rangle$ can be regarded as an improved version of $\mathcal{E}$
that takes into account the disc velocity dispersion, which is an important
ingredient missing from $\mathcal{E}$ (Athanassoula 2008).  Note also that
$\langle Q\rangle$ can easily be corrected so as to take into account the
vertical structure of the disc (Romeo \& Mogotsi 2018), but that correction
cancels out in the final results (Romeo 2020).

\begin{figure*}
\includegraphics[scale=1.26]{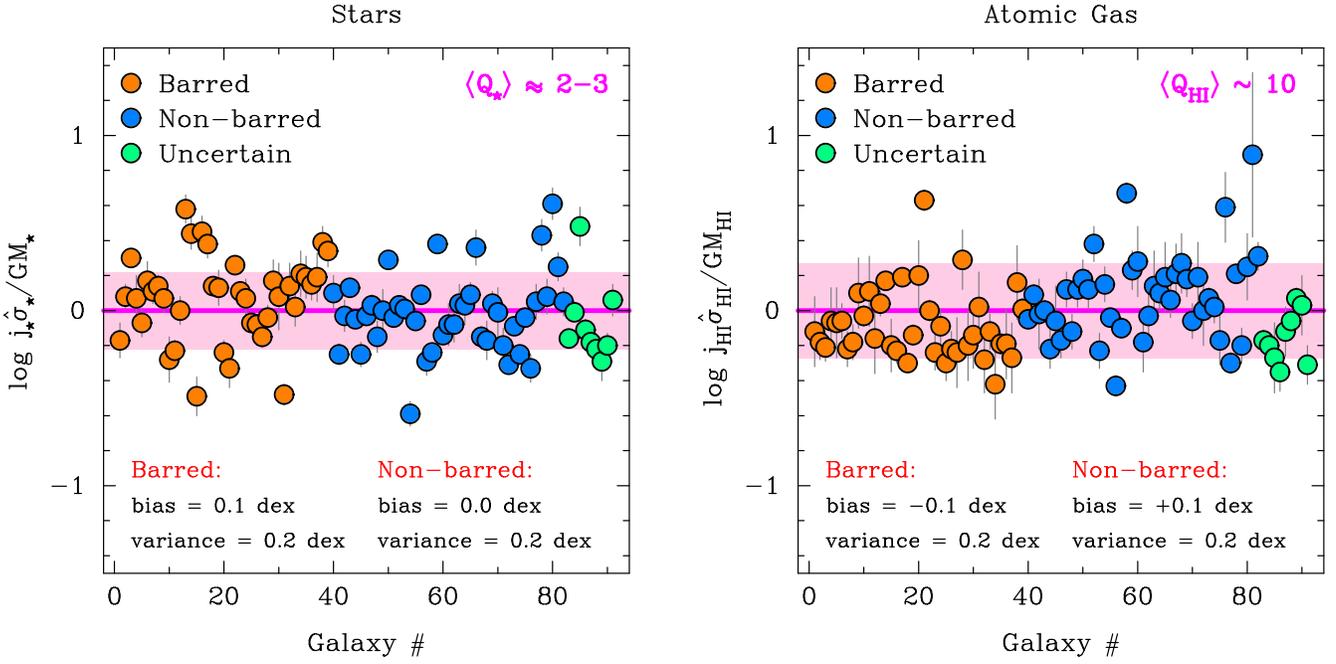}
\caption{Self-regulation of galaxy discs driven by local gravitational
  instabilities: barred versus non-barred galaxies.  The galaxy sample and
  the data are described in Sect.\ 2.  The magenta lines are the
  parameter-free theoretical predictions made by Romeo (2020) for stars and
  atomic gas,
  $j_{i}\hat{\sigma}_{i}/GM_{i}\approx1\;(i=\star,\mbox{H\,\textsc{i}})$,
  where this quantity is a normalized proxy for the mass-weighted average of
  Toomre's (1964) $Q_{i}$ stability parameter (see Sect.\ 4.2.2 for more
  information).  The pink regions are the observed $1\sigma$ scatters.  The
  robust median-based statistics shown in the left and right panels disclose
  a weak effect, which is also visually detectable as a small vertical offset
  between the orange and azure data points: barred galaxies self-regulate to
  systematically larger values of $\langle Q_{\star}\rangle$ and smaller
  values of $\langle Q_{\mathrm{HI}}\rangle$ than non-barred galaxies.  Once
  such biases are taken into account, both types of galaxies exhibit the same
  cosmic variance in $\langle Q\rangle$: 0.2 dex, a universal value for both
  stars and atomic gas.}
\end{figure*}

\begin{figure*}
\includegraphics[scale=1.26]{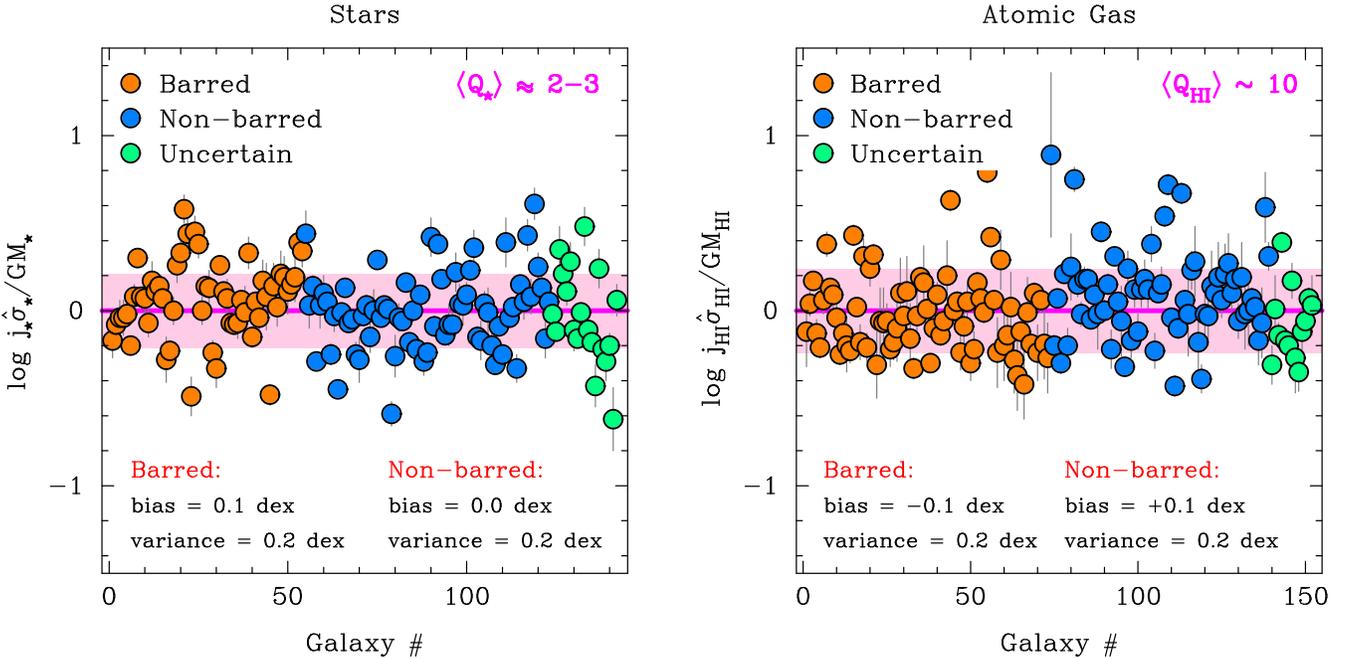}
\caption{Self-regulation of galaxy discs driven by local gravitational
  instabilities: robustness of the results.  This is similar to Fig.\ 4, but
  here the left and right panels show data from two of the largest samples of
  galaxies with quality-assessed measurements of $M_{\star}$, $j_{\star}$
  (142 galaxies) and $M_{\mathrm{HI}}$, $j_{\mathrm{HI}}$ (152 galaxies),
  respectively.  See item (i) of Sect.\ 4.2.2 for more information.}
\end{figure*}

\begin{figure*}
\includegraphics[scale=1.26]{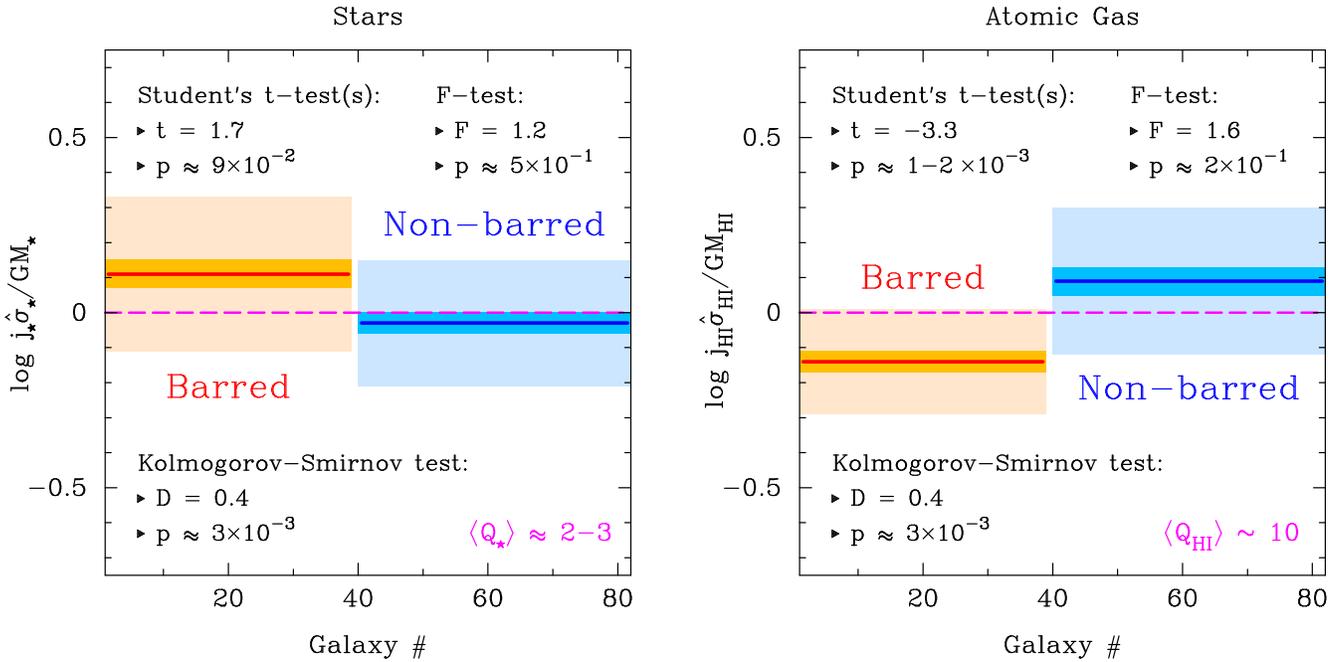}
\caption{Self-regulation of galaxy discs driven by local gravitational
  instabilities: statistical validity and significance of the results.  The
  galaxy sample is the same as in Fig.\ 4, but here the barred and non-barred
  data sets are analysed using a variety of statistical diagnostics
  (`uncertain' galaxies are not considered).  Each panel shows the median
  values (solid lines), robust standard errors (thin rectangles) and robust
  standard deviations (thick rectangles) of
  $\log\,j_{i}\hat{\sigma}_{i}/GM_{i}\;(i=\star,\mbox{H\,\textsc{i}})$ for
  the two data sets, together with the predicted value of $\langle
  Q_{i}\rangle$ (Romeo 2020).  Also shown are several comparative tests.  See
  item (ii) of Sect.\ 4.2.2 for more information.}
\end{figure*}

\subsubsection{Do bars have any impact on $\langle Q_{\star}\rangle$ or
               $\langle Q_{\mathrm{HI}}\rangle$?}

An aspect of the angular momentum problem that connects bar structure in
galaxies with the self-regulation of galaxy discs concerns the impact that
bars may have on the characteristic values of $\langle Q\rangle$ discussed
above.  This clearly deserves to be explored, since bars are well-known
drivers of secular evolution in disc galaxies (see, e.g., Gadotti 2009;
Combes 2011; Athanassoula 2013; Kormendy 2013).  In fact, bars are not rigid
structures that possess a fixed amount of energy and angular momentum.  Bars
are complex adaptive systems that grow and sustain themselves at the expense
of the gravitational potential energy of the galaxy, and that transfer
angular momentum to the outer disc and to the halo during the whole process
of bar formation and evolution.  Spiral arms also transfer angular momentum
to the outer disc, but they do it less efficiently than bars (see references
above).

Our analysis is based on Romeo's (2020) key equation, which naturally
connects with the analysis carried out in previous sections:
\begin{equation}
\frac{j_{i}\hat{\sigma}_{i}}{GM_{i}}\approx1\;\;\;\;\;
\mbox{for\ }i=\star,\,\mbox{H\,\textsc{i}},\,\mbox{H}_{2}\,.
\end{equation}
This a tight statistical relation between mass ($M$), specific angular
momentum ($j$) and velocity dispersion ($\hat{\sigma}$) for each baryonic
component in the disc plus bulge: stars ($i=\star$), atomic hydrogen + helium
gas ($i=\mbox{H\,\textsc{i}}$) and molecular hydrogen + helium gas
($i=\mbox{H}_{2}$).  To make good use of Eq.\ (11), one needs to understand
two key points:
\begin{itemize}
\item $j_{i}\hat{\sigma}_{i}/GM_{i}$ is a normalized proxy for $\langle
  Q_{i}\rangle$, which itself is more difficult to evaluate accurately and
  approximate analytically.  The normalization is such that
  $j_{i}\hat{\sigma}_{i}/GM_{i}\approx1$ corresponds to $\langle
  Q_{\star}\rangle\approx2\mbox{--}3$, $\langle Q_{\mathrm{HI}}\rangle\sim10$
  and $\langle Q_{\mathrm{H2}}\rangle\sim10$.  These values are
  parameter-free theoretical predictions that have an expected accuracy of
  about 0.2 dex (see sect.\ 2 of Romeo 2020).  We will make use of such
  predictions when presenting the results of our analysis, so as to translate
  from $j_{i}\hat{\sigma}_{i}/GM_{i}$ into $\langle Q_{i}\rangle$.
\item $\hat{\sigma}_{i}$ is the radial velocity dispersion of component $i$,
  $\sigma_{i}$, properly averaged and rescaled.  This quantity can be
  evaluated using two alternative equations, depending on whether there are
  reliable $\sigma_{i}$ measurements available or not.  Unfortunately, such
  measurements are highly non-trivial (e.g., Ianjamasimanana et al.\ 2017;
  Marchuk \& Sotnikova 2017), hence very sparse (e.g., Romeo \& Mogotsi 2017;
  Mogotsi \& Romeo 2019).  Therefore, if one wants to analyse a large galaxy
  sample, then the appropriate equation to use is
  \begin{equation}
  \hat{\sigma}_{i}\approx
  \left\{
  \begin{array}{rll}
  130\;\mbox{km\,s}^{-1}
  & \!\!\!\!\times\,\,(M_{\star}/10^{10.6}\,\mbox{M}_{\odot})^{0.5}
  & \mbox{if\ }i=\star\,, \\
   11\;\mbox{km\,s}^{-1}
  &
  & \mbox{if\ }i=\mbox{H\,\textsc{i}}\,, \\
    8\;\mbox{km\,s}^{-1}
  &
  & \mbox{if\ }i=\mbox{H}_{2}\,.
  \end{array}
  \right.
  \end{equation}
  Note that these are not observationally motivated values of the stellar and
  gas velocity dispersions, but rigorously derived values of the velocity
  dispersion--based quantity $\hat{\sigma}_{i}$ (see sect.\ 2 of Romeo 2020).
  Our analysis is based on both Eq.\ (11) and Eq.\ (12).
\end{itemize}

Eq.\ (11), when combined with Eq.\ (12), shows a statistical scatter of about
0.2 dex (Romeo 2020; Romeo et al.\ 2020), which hides a slight systematic
trend with gas mass fraction (see fig.\ 3 of Romeo et al.\ 2020), also
detected by another careful analysis (Mancera Pi\~{n}a et al.\ 2021b).  That
trend does not affect the usefulness of Eqs (11) and (12), which has been
highlighted by Romeo (2020) and Romeo et al.\ (2020), and confirmed by
independent investigations (e.g., Kurapati et al.\ 2021; Bouch\'{e} et
al.\ 2022).  Hereafter we will make use of such equations without considering
molecular gas, since CO data are not available for most galaxies of our
sample, as already mentioned in Sect.\ 2.2.

Let us now analyse in detail how barred/non-barred disc galaxies
self-regulate their stellar and atomic gas components via local gravitational
instabilities.  The first part of our analysis is illustrated in Fig.\ 4,
where each panel shows the predicted value of $j_{i}\hat{\sigma}_{i}/GM_{i}$
(magenta line), the corresponding approximate value of $\langle
Q_{i}\rangle$, the observed $1\sigma$ scatter (pink region), as well as
statistical information about the barred and non-barred data sets: their
`bias' (median offset from the prediction), and their `variance' (robust
standard deviation from the median trend).  Such statistical diagnostics
disclose a weak systematic effect, which is also visually detectable as a
small vertical offset between the orange and azure data points: barred
galaxies self-regulate to values of $\langle Q_{\mathrm{HI}}\rangle$ (values
of $\langle Q_{\star}\rangle$) that are typically $\approx$ 0.2 dex smaller
($\approx$ 0.1 dex larger) than those of non-barred galaxies.  In contrast,
both types of galaxies exhibit the same cosmic variance in $\langle
Q\rangle$: $\approx$ 0.2 dex, a universal value for both stars and atomic
gas.

The results presented above are reliable because they are based on robust
statistics, and because the barred/non-barred data sets are statistically
unbiased (see Sect.\ 2.2 for more information).  Furthermore, such results
are new and unexpected.  Indeed, we expected to find a clearer separation
between barred and non-barred galaxies than that shown in Fig.\ 4, given that
bars are expected to enrich the outer disc with a significant amount of
angular momentum (see the opening paragraph of this section), and given that
the outskirts of the disc contribute most to $j_{\star}$ and
$j_{\mathrm{HI}}$.  Below we verify such results carefully.
\begin{enumerate}
\item \textbf{Robustness of the results.}  Let us re-analyse the effect of
  bars on $\langle Q_{\star}\rangle$ and $\langle Q_{\mathrm{HI}}\rangle$
  using two larger data sets: $\{M_{\star},j_{\star}\}$ from a sample of 142
  galaxies selected by Romeo et al.\ (2020), and
  $\{M_{\mathrm{HI}},j_{\mathrm{HI}}\}$ from a sample of 152 galaxies
  selected by Mancera Pi\~{n}a et al.\ (2021a).  These are two of the largest
  galaxy samples with quality-assessed measurements of such quantities.
  Remember that their intersection is the galaxy sample described in
  Sect.\ 2.  This second part of our analysis is illustrated in Fig.\ 5.  The
  larger sample size makes the distribution of data points denser inside the
  pink regions, hence statistically closer to the magenta lines, thus
  highlighting how well self-regulated disc galaxies are.  The larger sample
  size also makes it clearer that barred galaxies are characterized by median
  values of $\langle Q_{\star}\rangle$ and $\langle Q_{\mathrm{HI}}\rangle$
  that are different from those of non-barred galaxies, while both types of
  galaxies have the same cosmic variance in $\langle Q\rangle$.
\item \textbf{Statistical validity and significance of the results.}  Let us
  finally re-consider the galaxy sample described in Sect.\ 2 and analyse the
  barred/non-barred data sets in detail using a variety of statistical
  diagnostics (`uncertain' galaxies are not considered).  This third part of
  our analysis is illustrated in Fig.\ 6, where each panel shows the median
  values (solid lines), robust standard errors (thin rectangles) and robust
  standard deviations (thick rectangles) of
  $\log\,j_{i}\hat{\sigma}_{i}/GM_{i}$ for the two data sets, as well as
  several comparative tests.  These general statistical tests are described
  in chaps 14.2 and 14.3 of Press et al.\ (1992), for example, and quantify
  how significantly different two distributions are as regards their means
  (Student's t-test), variances (F-test) and cumulative behaviours, most
  sensitively around the median values (Kolmogorov-Smirnov test).  All such
  diagnostics speak clearly: bars have a weak but significant impact on
  $\langle Q_{\mathrm{HI}}\rangle$ and an opposite feeble effect on $\langle
  Q_{\star}\rangle$, which is at the limit of statistical significance (see
  Fig.\ 6 for detailed information).
\end{enumerate}

\section{CONCLUSIONS}

In this paper, we have used publicly available measurements of mass ($M$) and
specific angular momentum ($j$) of various galaxy components, which result
from the work of several teams (Read et al.\ 2017; Posti et al.\ 2018, 2019;
Mancera Pi\~{n}a et al.\ 2021a).  Using such high-quality data with a wide
dynamic range from the SPARC and the LITTLE THINGS galaxy samples, we have
performed a detailed comparative analysis that stretches across a variety of
galaxy properties, for disc galaxies of all morphological types: from
lenticulars to blue compact dwarfs.  Our analysis solves important aspects of
the angular momentum problem, and imposes tight constraints not only on $j$
itself but also on its connection with galaxy morphology, bar structure and
disc gravitational instability.  Our major results are pointed out below.
\begin{itemize}
\item The `baryons' (stars and atomic gas) contained in the discs and bulges
  of nearby star-forming galaxies have retained, globally, slightly more than
  80\% of the specific angular momentum possessed by their host dark matter
  haloes.  Thus $j$ is conserved to better than 20\% in the process of disc
  galaxy formation and evolution (see Fig.\ A1 for detailed statistical
  information).  There is however a clear systematic trend: the retained
  fraction of specific angular momentum, $j_{\mathrm{b}}/j_{\mathrm{h}}$,
  depends on the galaxy formation efficiency,
  $M_{\mathrm{b}}/M_{\mathrm{h}}$, and varies on average as
  $j_{\mathrm{b}}/j_{\mathrm{h}}\propto
  (M_{\mathrm{b}}/M_{\mathrm{h}})^{0.5}$.  This correlation is moderately
  strong but very significant (e.g.\ Spearman's $\rho\approx0.5$ and
  $p_{\rho}\sim10^{-7}$).  In contrast, $j_{\mathrm{b}}/j_{\mathrm{h}}$ does
  not show any particularly significant ($p_{\rho}\la10^{-4}$) correlation
  with basic galaxy properties like $j_{\mathrm{h}}$, $M_{\mathrm{h}}$ or
  their baryonic counterparts.
\item Stars have about 40\% \emph{less} specific angular momentum than the
  halo, whereas atomic gas has about 20\% \emph{more} (see Fig.\ A1 for
  detailed statistical information).  This implies that
  $j_{\mathrm{HI}}\approx2j_{\star}$, which confirms a result previously
  found by Mancera Pi\~{n}a et al.\ (2021b).  There is a clear systematic
  trend even for these two baryonic components: $j_{i}/j_{\mathrm{h}}\propto
  (M_{i}/M_{\mathrm{h}})^{A_{i}}$.  Stars show a gentler logarithmic slope
  ($A_{\star}=0.2$) than atomic gas ($A_{\mathrm{HI}}=0.5$), but a comparable
  degree of correlation ($\rho\approx0.4$ and
  $p_{\rho}\sim10^{-4}\mbox{--}10^{-5}$).  The two scaling relations
  $j_{i}/j_{\mathrm{h}}$ vs $M_{i}/M_{\mathrm{h}}$ show opposite residual
  trends with galaxy morphology.  Early-type disc galaxies tend to cluster
  below (above) the best-fitting relation found for stars (atomic gas), hence
  they tend to have lower $j_{\star}/j_{\mathrm{h}}$ (higher
  $j_{\mathrm{HI}}/j_{\mathrm{h}}$) than predicted.  This tendency is
  reversed for late-type disc galaxies.  As in the case of baryons, we have
  also analysed the relations between $j_{i}/j_{\mathrm{h}}$ and basic galaxy
  properties, and found that stars and atomic gas show similarities as well
  as significant differences (see Appendix A for more information).
\item The fraction of specific angular momentum retained by the disc,
  $j_{\mathrm{d}}/j_{\mathrm{h}}$, is one of the building blocks of the ELN
  bar instability criterion.  This simple criterion, which is used by all
  current semi-analytic models of galaxy formation and evolution, is believed
  to be less inaccurate than other uncertainties of the modelling, although
  careful simulation tests suggest otherwise (see Sect.\ 4.1.1 for an
  overview).  Our observational test, which is the first of its kind and is
  based on a statistically unbiased sample of barred/non-barred galaxies,
  demonstrates that the ELN criterion is highly inaccurate: it fails in about
  55\% of the cases.  A more fundamental limitation is that barred and
  non-barred galaxies are mixed across the entire range of values spanned by
  the ELN parameter, $\mathcal{E}$, and cannot thus be separated by shifting
  the instability threshold ($\approx1$) up or down, regardless of whether
  the ELN criterion is applied to the whole baryonic disc or to its stellar
  component.  Our test further demonstrates that the ELN criterion tends to
  overpredict how bar stable disc galaxies are, especially in the case of
  low-mass gas-rich spirals and dwarfs.  Such a tendency is exacerbated when
  one uses the popular stellar-disc implementation of the ELN criterion.
\item $j_{\star}$ and $j_{\mathrm{HI}}$ enter another important galaxy
  evolution process, which takes place in disc galaxies of all morphological
  types: the self-regulation of galaxy discs driven by local gravitational
  instabilities (see Sect.\ 4.2.1 for an overview).  Using a variety of
  statistical diagnostics, we have shown that bars have a weak but
  significant impact on such a process: barred galaxies self-regulate to
  values of $\langle Q_{\mathrm{HI}}\rangle$ that are typically $\approx$ 0.2
  dex smaller than those of non-barred galaxies, where $\langle
  Q_{\mathrm{HI}}\rangle$ is the mass-weighted average of Toomre's (1964)
  $Q_{\mathrm{HI}}$ stability parameter.  We have also detected an opposite,
  $\approx$ 0.1 dex effect on $\langle Q_{\star}\rangle$, but the signal is
  so faint that this effect is at the limit of statistical significance.
  Despite these systematic trends, both barred and non-barred galaxies
  exhibit the same cosmic variance in $\langle Q\rangle$: $\approx$ 0.2 dex,
  a universal value for both stars and atomic gas.
\end{itemize}

Our results on barred galaxies are of particular interest for semi-analytic
modelling of galaxy formation and evolution.  Thus we want to clarify them
further, and highlight the differences between $\mathcal{E}$ and $\langle
Q\rangle$.

First of all, it is amazingly challenging to characterize barred galaxies
from a gravitational instability point of view.  In the best of the cases,
the signal is faint and appropriate statistical methods are required to
separate it from the noise.  \emph{This is an important point to keep in
  mind!}

Secondly, it is unexpected but not totally surprising that $\mathcal{E}$
shows no signal at all, while $\langle Q\rangle$ shows a faint signal.  On
the one hand, no parameter can represent the complex phenomenology of bars in
disc galaxies or the disc-halo interaction, which is vital for bars
(Athanassoula 2008; Sellwood 2016).  On the other hand, $\langle Q\rangle$
takes into account the disc velocity dispersion, $\langle
Q\rangle\propto\mathcal{E}^{2}\,\sigma/V$ (Romeo \& Mogotsi 2018), which is
an important ingredient missing from $\mathcal{E}$ (Athanassoula 2008).  It
may still seem strange that a quantity derived from the local stability
parameter $Q$ `feels' the presence of bars, which are classically associated
with global gravitational instability.  But bars produce redistribution of
matter in the disc, which alters the radial profile of $Q$ (Romeo \& Fathi
2015, 2016), hence $\langle Q\rangle$.

Last but not least, \emph{neither $\langle Q_{\star}\rangle$ nor $\langle
  Q_{\mathrm{HI}}\rangle$ is a bar instability parameter, and neither of them
  should be used as such!}  Use instead $\langle Q_{\star}\rangle$, and
especially $\langle Q_{\mathrm{HI}}\rangle$, to test whether
modelled/simulated barred galaxies behave like the observed ones.  $\langle
Q_{\mathrm{H2}}\rangle$ is potentially a more useful diagnostic than $\langle
Q_{\mathrm{HI}}\rangle$.  This is suggested by the fact that bars can host
large amounts of molecular gas (Renaud et al.\ 2015), and by the fact that
molecular gas does not extend so far out in the disc as atomic gas, hence it
is a more sensitive tracer of the bar gravitational potential.
Unfortunately, there are no CO data available for most galaxies of our
sample, so we have not tested the ability of $\langle Q_{\mathrm{H2}}\rangle$
to distinguish barred from non-barred galaxies.  We leave that for future
work.

\section*{ACKNOWLEDGEMENTS}

ABR dedicates this paper to his wife {\AA}sa: for your immense love and
support.  We are very grateful to Martin Fr\"{o}st for musical inspiration,
to Pavel Mancera Pi\~{n}a for help with the data, and to Dimitrios Irodotou,
David Izquierdo-Villalba, Pavel Mancera Pi\~{n}a and Robert Nau for useful
discussions.  We are also very grateful to an anonymous referee for
insightful comments and suggestions, and for encouraging future work on the
topic.  OA and FR acknowledge support from the Knut and Alice Wallenberg
Foundation.  OA acknowledges support from the Swedish Research Council (grant
2019-04659).

\section*{DATA AVAILABILITY}

The data underlying this article will be shared on reasonable request to the
corresponding author.

\appendix

\section{ADDITIONAL FIGURES}

This appendix contains four additional figures.

Fig.\ A1, mentioned in Sects 3.1 and 3.2, provides detailed statistical
information concerning $\log\,j_{\mathrm{b}}/j_{\mathrm{h}}$,
$\log\,j_{\star}/j_{\mathrm{h}}$ and $\log\,j_{\mathrm{HI}}/j_{\mathrm{h}}$.
The most important point illustrated by this figure is that the probability
distributions of such fractions have a strong central tendency.  This is
especially true for baryons and stars, whose distributions are clearly
unimodal and more peaked than a Gaussian.  In such cases, the median is a
robust estimator of the central value of the distribution (see chap.\ 14.1 of
Press et al.\ 1992).  Thus the median values $\pm$ robust standard errors of
$j_{\mathrm{b}}/j_{\mathrm{h}}$, $j_{\star}/j_{\mathrm{h}}$ and
$j_{\mathrm{HI}}/j_{\mathrm{h}}$ provide fully meaningful estimates of how
well specific angular momentum is conserved in a statistical sense,
regardless of how strongly or significantly the retained fractions of $j$
correlate with other galaxy properties.

Figs A2--A4, also mentioned in Sects 3.1 and 3.2, supplement the information
provided by Figs 1 and 2 with additional correlation plots.  Figs A2 and A3
show that $j_{\mathrm{b}}/j_{\mathrm{h}}$ and $j_{\star}/j_{\mathrm{h}}$ do
not have any particularly significant ($p\la10^{-4}$) correlation with basic
galaxy properties like $j_{\mathrm{h}}$, $M_{\mathrm{h}}$ or their
baryonic/stellar counterparts.  Fig.\ A4 shows that
$j_{\mathrm{HI}}/j_{\mathrm{h}}$ does not correlate with $j_{\mathrm{h}}$ or
$M_{\mathrm{h}}$, while it has a moderately high degree of correlation with
$j_{\mathrm{HI}}$ and $M_{\mathrm{HI}}$.

\begin{figure*}
\includegraphics[scale=0.95]{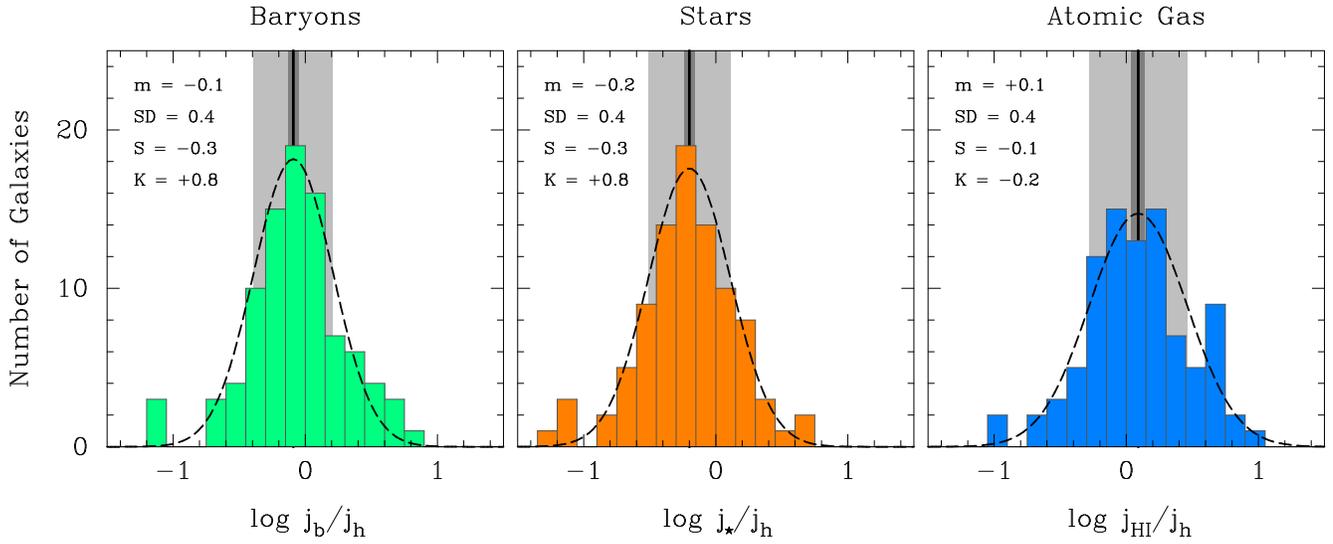}
\caption{Detailed statistical information concerning
  $\log\,j_{\mathrm{b}}/j_{\mathrm{h}}$, $\log\,j_{\star}/j_{\mathrm{h}}$ and
  $\log\,j_{\mathrm{HI}}/j_{\mathrm{h}}$, the logarithmic fractions of
  specific angular momentum retained by baryons, stars and atomic gas.  The
  galaxy sample and the data are described in Sect.\ 2.  Each panel shows the
  observed probability distribution (histogram) together with several robust
  statistics: the median, $\mbox{Med}$ (solid line), the robust standard
  error, $\mbox{SE}_{\mathrm{rob}}$ (narrow stripe), the robust standard
  deviation, $\mbox{SD}_{\mathrm{rob}}$ (wide stripe), as well as a Gaussian
  probability distribution with parameters $\mu=\mbox{Med}$ and
  $\sigma=\mbox{SD}_{\mathrm{rob}}$ normalized as the histogram (dashed
  curve).  Also reported are the values of several classical statistics: the
  mean ($m$), standard deviation ($\mbox{SD}$), skewness ($S$) and kurtosis
  ($K$).  Positive/negative values of $S$ mean that the probability
  distribution has a longer tail on the right/left.  Positive/negative values
  of $K$ mean that the probability distribution has fatter/thinner tails than
  a Gaussian distribution, which often implies that the distribution is more
  peaked/flat than a Gaussian.}
\end{figure*}

\begin{figure*}
\includegraphics[scale=1.13]{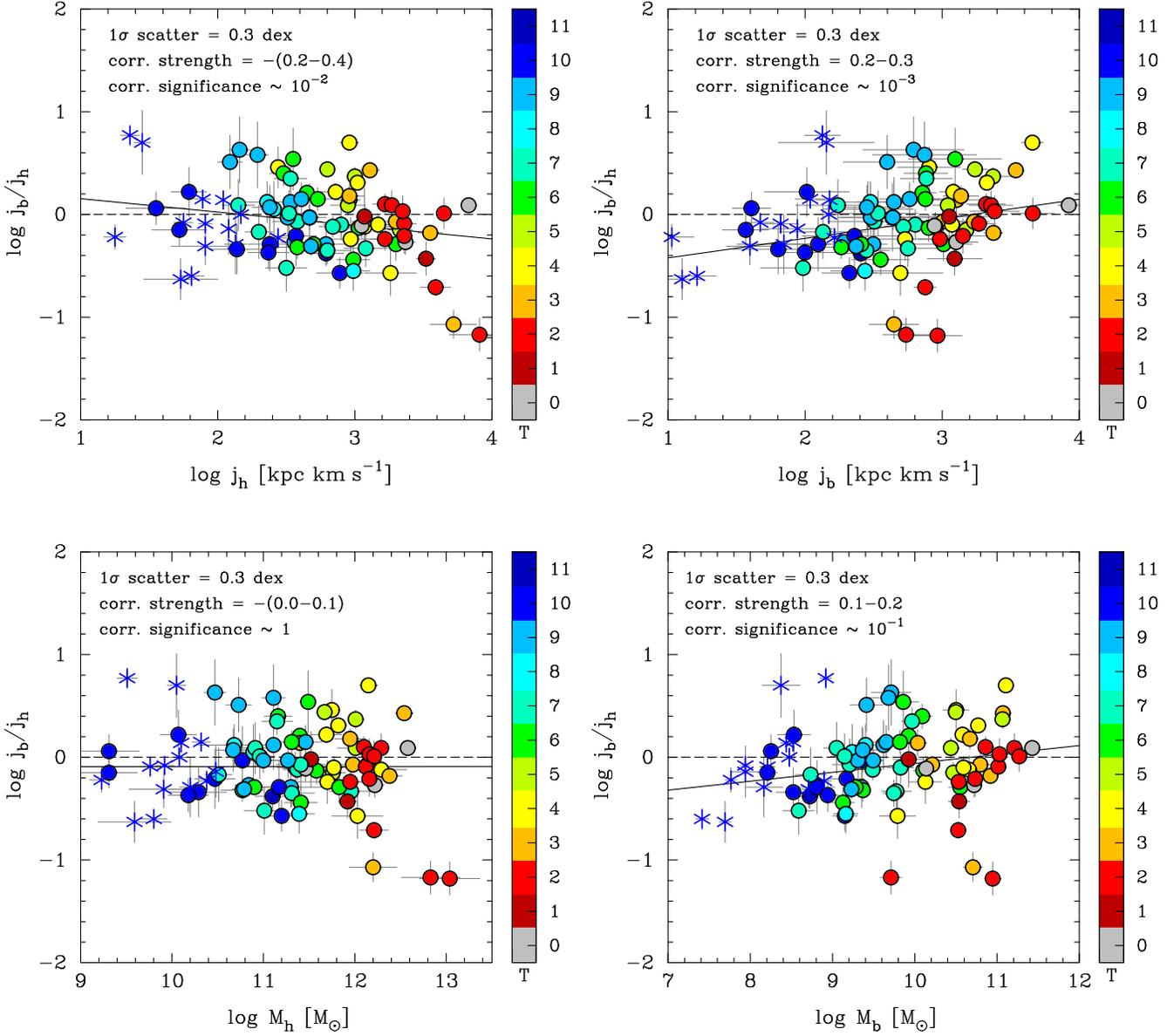}
\caption{The fraction of specific angular momentum retained by baryons
  ($j_{\mathrm{b}}/j_{\mathrm{h}}$) versus basic galaxy properties.  These
  correlations are weaker and less significant than those shown in Fig.\ 1.
  The galaxy sample and the data are described in Sect.\ 2.  Galaxies are
  colour-coded by Hubble stage, and symbol-coded by their parent samples:
  SPARC (solid circles with black ouline) and LITTLE THINGS (asterisks).  The
  solid lines are robust median-based fits to the data points (see Sect.\ 2.3
  for more information).  The dashed lines indicate conservation of specific
  angular momentum, i.e.\ that baryons have retained the same amount of
  specific angular momentum as the host dark matter halo.  Statistical
  information about the data is given in summary form and simplified notation
  (see Sect.\ 2.3 for more information).}
\end{figure*}

\begin{figure*}
\includegraphics[scale=1.13]{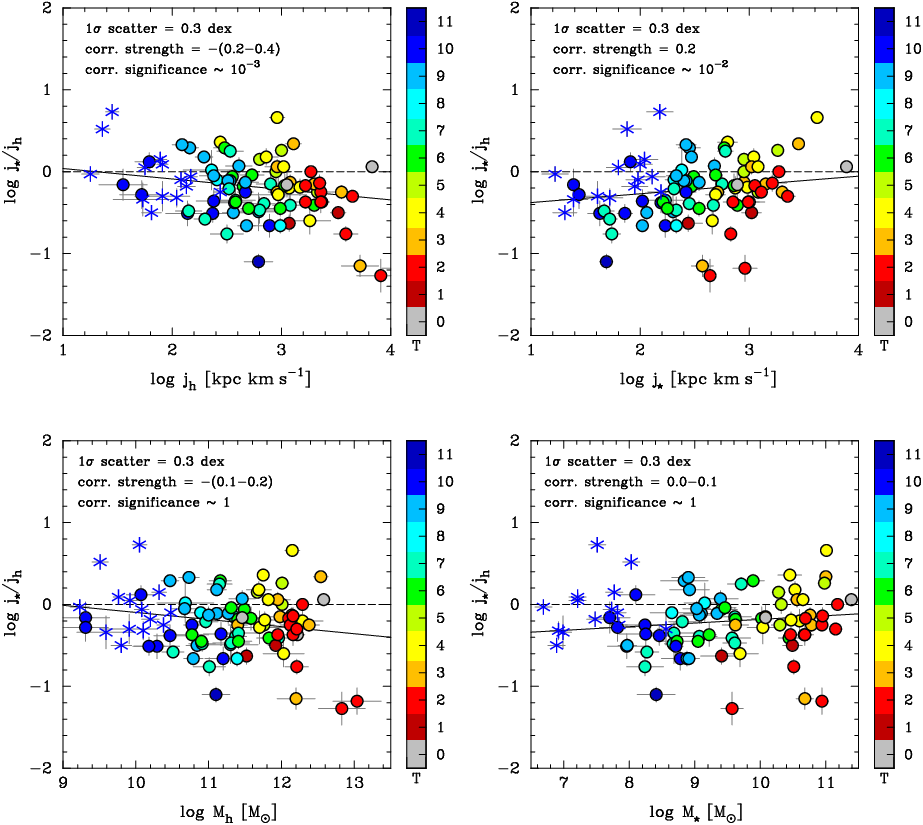}
\caption{Same as Fig.\ A2, but for the fraction of specific angular momentum
  retained by the stellar component ($j_{\star}/j_{\mathrm{h}}$).}
\end{figure*}

\begin{figure*}
\includegraphics[scale=1.13]{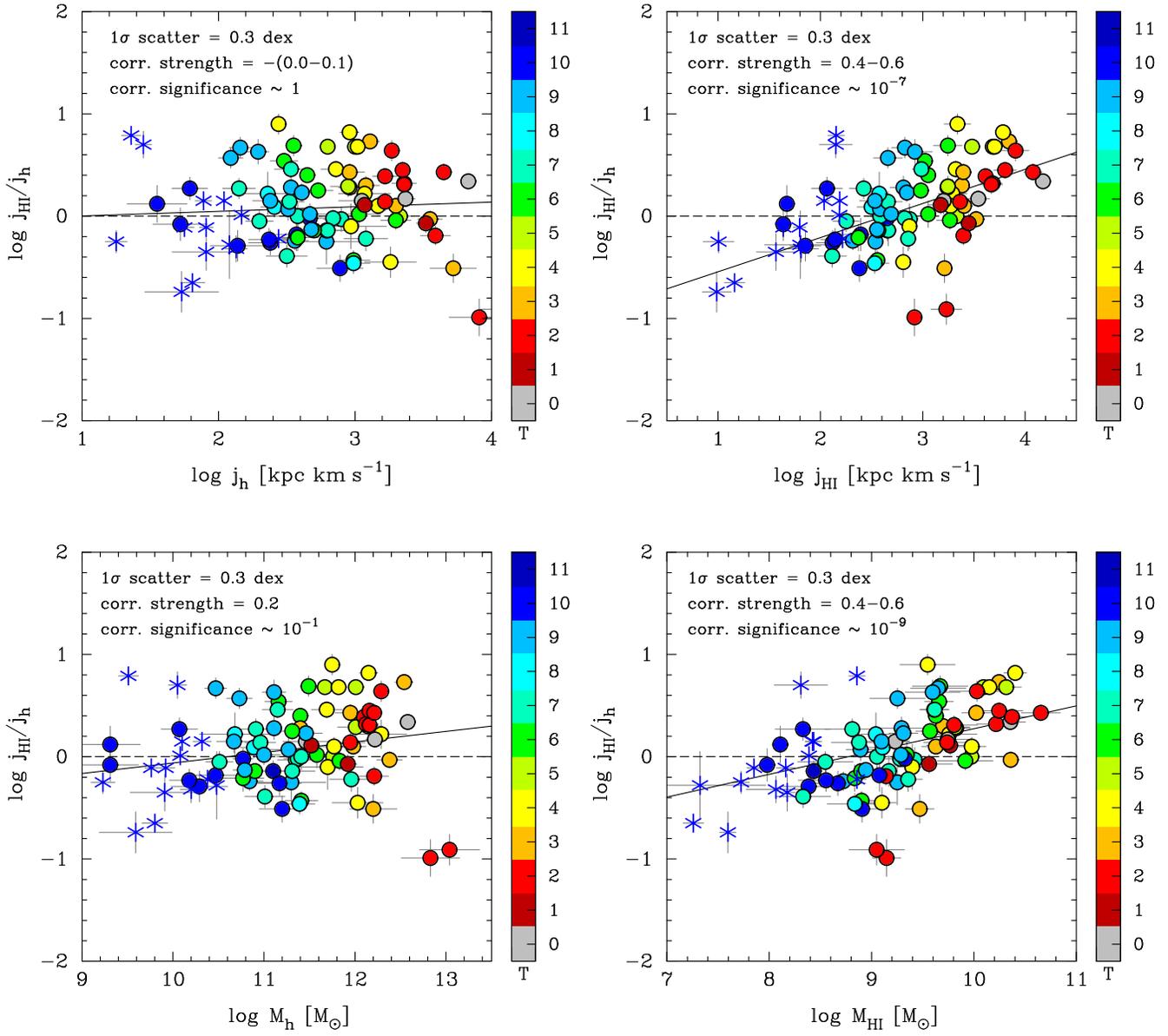}
\caption{Same as Fig.\ A2, but for the fraction of specific angular momentum
  retained by the atomic gas component ($j_{\mathrm{HI}}/j_{\mathrm{h}}$).}
\end{figure*}

\section{ADDITIONAL TESTS}

This appendix describes three additional tests.

The first test, mentioned in Sect.\ 2.2, probes two numerical aspects of the
problem: (i) the effect of varying the random realization of $\lambda$ on the
correlations between $j_{\mathrm{h}}$, $M_{\mathrm{h}}$ and $M_{\mathrm{b}}$;
and (ii) the impact of suppressing the natural variance of $\lambda$ on such
correlations.  To probe (i), we run 10 Monte Carlo simulations of
$j_{\mathrm{h}}=j_{\mathrm{h}}(\lambda)$ for our galaxy sample, i.e.\ we
randomly generate 10 sets of 91 values of $\lambda$ from Eq.\ (3), setting
$\lambda_{0}=0.035$ and $\sigma=0.50$ (0.22 dex) as in Sect.\ 2.2, and
compute $j_{\mathrm{h}}$ from Eq.\ (4).  For each simulation, we analyse
$j_{\mathrm{h}}$ vs $M_{\mathrm{h}}$ and $j_{\mathrm{h}}$ vs $M_{\mathrm{b}}$
using the statistics described in Sect.\ 2.3.  We then evaluate the mean and
the standard deviation of each statistic over the ensemble of simulations.
To probe (ii), we set $\lambda=0.035$ in Eq.\ (4), and analyse the
$j_{\mathrm{h}}\mbox{--}M_{\mathrm{h}}$ and
$j_{\mathrm{h}}\mbox{--}M_{\mathrm{b}}$ relations using our statistical
toolkit.  Fig.\ B1 illustrates all such information, including the fiducial
random realization set used in this paper, which is highlighted in orange
over a `sea' of azure data points.  Our test demonstrates that varying the
random realization of $\lambda$ has a weak ($\la10\%$) effect on the results,
whereas suppressing the natural variance of $\lambda$ artificially constrains
the correlations between $j_{\mathrm{h}}$ and other fundamental galaxy
properties like $M_{\mathrm{h}}$ and $M_{\mathrm{b}}$.

The second test, mentioned in Sect.\ 4.1.2, checks whether the low accuracy
found for the ELN criterion is an artefact of using the reformulation made by
Mo et al.\ (1998) [Eq.\ (10)], rather than the original criterion (Efstathiou
et al.\ 1982) [Eq.\ (9)].  To check this, we need to evaluate two additional
quantities: the exponential disc scale length, $R_{\mathrm{d}}$, and the
maximum rotation velocity, $V_{\mathrm{max}}$.  Since there are no publicly
available measurements of $V_{\mathrm{max}}$ for most galaxies of our sample,
we use $V_{\mathrm{flat}}$ as a proxy for $V_{\mathrm{max}}$, where
$V_{\mathrm{flat}}$ is the velocity along the flat part of the rotation
curve.  $R_{\mathrm{d}}$ and $V_{\mathrm{flat}}$ are taken from Lelli et
al.\ (2016) for SPARC galaxies, and from Hunter \& Elmegreen (2006) and Iorio
et al.\ (2017) for LITTLE THINGS galaxies, respectively.  More precisely, for
8 of the 77 SPARC galaxies, $V_{\mathrm{flat}}$ is undefined because the
rotation curve does not reach a flat part.  Hence the galaxy sample used for
this test contains 83 galaxies in total.  Fig.\ B2 shows that the low
accuracy found in Sect.\ 4.1.2 is not an artefact of using the ELN criterion
reformulated by Mo et al.\ (1998), rather than the original ELN criterion
itself: Eqs (9) and (10) are almost indistinguishable from a statistical
point of view.

The third test, also mentioned in Sect.\ 4.1.2, checks whether it is possible
to improve the accuracy of the ELN criterion by applying it to the atomic gas
disc, rather than to the stellar disc.  Fig.\ B3 shows that the accuracy of
the ELN criterion is low even in that case.

\begin{figure*}
\includegraphics[scale=1.25]{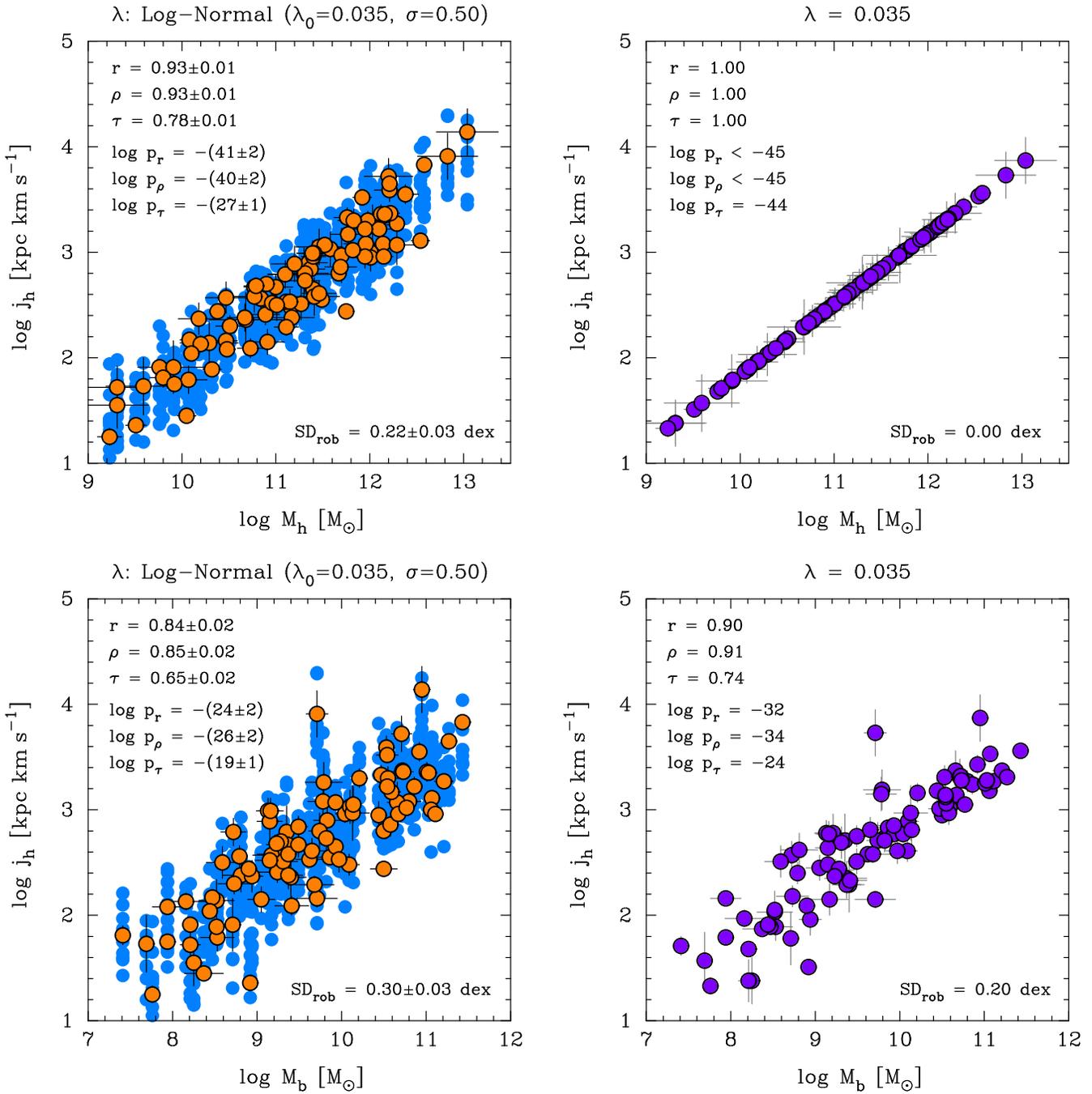}
\caption{Left panels: effect of varying the random realization of the halo
  spin parameter ($\lambda$) on the correlations between halo specific
  angular momentum ($j_{\mathrm{h}}$), halo mass ($M_{\mathrm{h}}$) and
  baryonic mass ($M_{\mathrm{b}}$).  Right panels: impact of suppressing the
  natural variance of $\lambda$ on such correlations.  The galaxy sample
  contains 91 galaxies and is described in Sect.\ 2, together with the data
  and the statistics.  The azure data points correspond to 10 sets of 91
  random realizations of $\lambda$, which are drawn from a log-normal
  probability distribution with median $\lambda_{0}=0.035$ and width
  $\sigma=0.50$ (0.22 dex).  The orange data points correspond to the
  fiducial random realization set used in this paper.  Statistical
  information shown in the left panels concerns the azure data points.  The
  value reported for each statistic is the mean $\pm$ standard deviation
  evaluated over the 10 random realization sets.  See Appendix B for more
  information.}
\end{figure*}

\begin{figure*}
\includegraphics[scale=1.26]{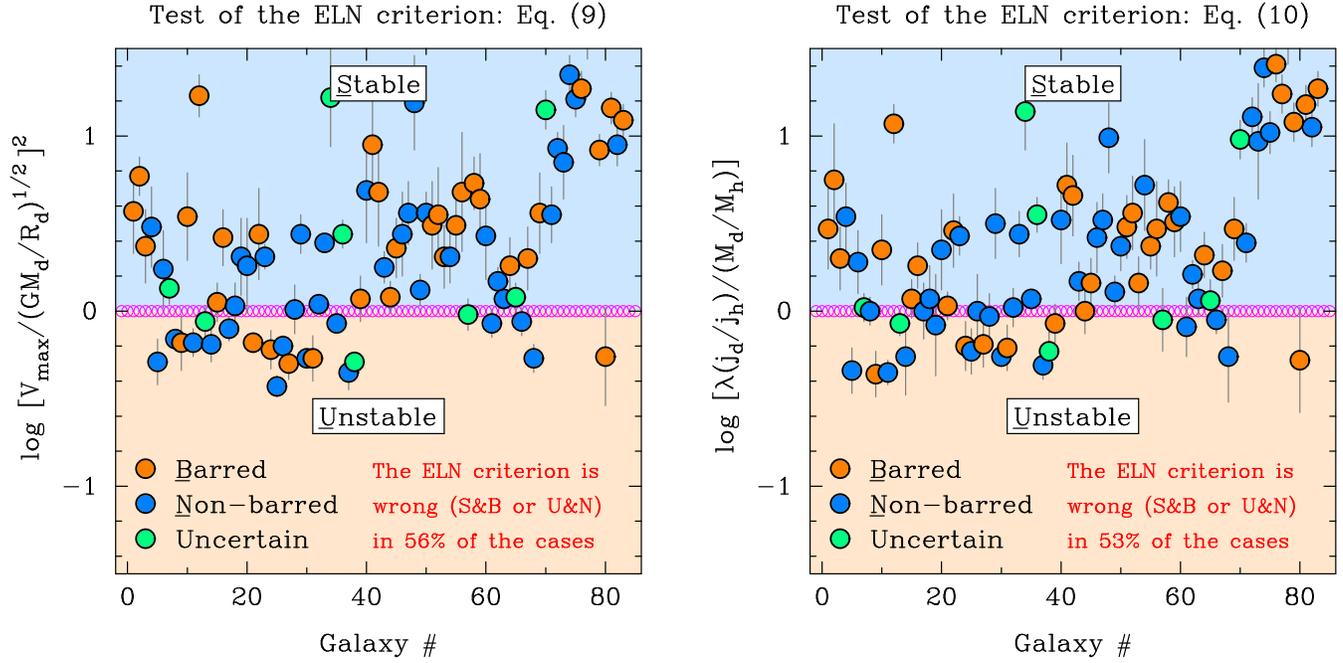}
\caption{Additional test of the ELN criterion: the original criterion
  (Efstathiou et al.\ 1982) [Eq.\ (9)] versus the reformulation made by Mo et
  al.\ (1998) [Eq.\ (10)].  See Appendix B for more information.}
\end{figure*}

\begin{figure*}
\includegraphics[scale=1.26]{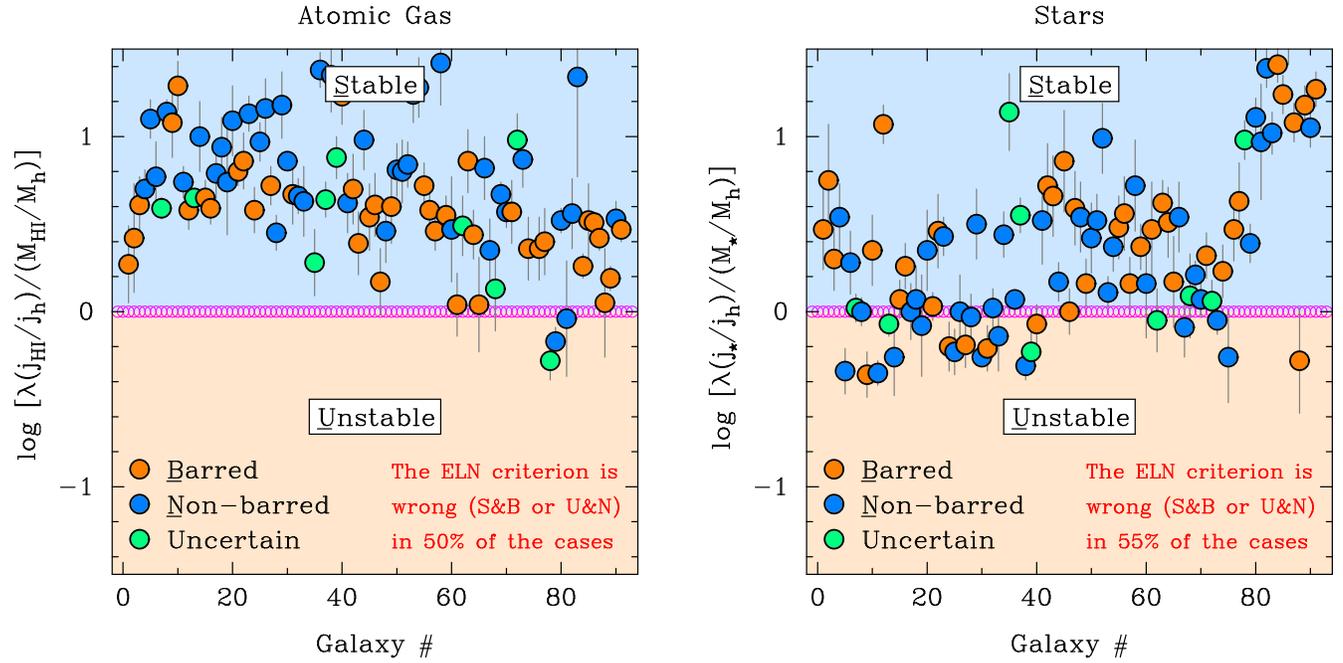}
\caption{Additional test of the ELN criterion: atomic gas versus stars.  See
  Appendix B for more information.}
\end{figure*}

\bsp

\label{lastpage}

\end{document}